\documentclass[journal]{IEEEtran}

\newcommand{\beq}{\begin{equation}}
\newcommand{\eeq}{\end{equation}}
\newcommand{\bsq}{\begin{subequations}}
\newcommand{\esq}{\end{subequations}}
\newcommand{\bq}{\begin{eqnarray}}
\newcommand{\eq}{\end{eqnarray}}
\newcommand{\bqn}{\begin{eqnarray*}}
\newcommand{\eqn}{\end{eqnarray*}}
\usepackage{bbm}
\usepackage[pdftex]{graphicx}
\graphicspath{{Figs/}}
\usepackage{enumerate}
\usepackage[hidelinks]{hyperref}

\usepackage{mathptmx}       
\DeclareMathAlphabet{\mathcal}{OMS}{cmsy}{m}{n}
\usepackage{helvet}         
\usepackage{courier}        
\usepackage{type1cm}        
\usepackage{makeidx}         
\usepackage{graphicx}        
\usepackage{url} 
\usepackage{multicol}        
\usepackage[bottom]{footmisc}
\usepackage{ifthen}
\usepackage{subfigure}
\usepackage[usenames,dvipsnames]{color}

\makeindex             

\usepackage{graphics} 
\usepackage{graphicx} 
\usepackage{epsfig} 
\usepackage{epstopdf}
\usepackage{mathptmx} 
\usepackage{times} 
\usepackage[cmex10]{amsmath}

\usepackage{mathtools}  
\usepackage{amssymb}  
\usepackage{amsthm}

\usepackage{amsfonts}
\usepackage{cite}
\usepackage{verbatim}
\usepackage{comment}
\usepackage{algorithm}   
\usepackage{algorithmic}
\usepackage{multirow}
\usepackage{cite}
\usepackage{stfloats}
\usepackage{supertabular}
\usepackage{longtable}

\DeclareGraphicsExtensions{.pdf,.png,.jpg,.eps,.ps}

\usepackage{arydshln}
\usepackage{multicol}
\usepackage{colortbl}
\theoremstyle{definition}

\newtheorem{proposition}{Proposition}

\theoremstyle{definition}
\newtheorem{definition}{Definition}

\ifCLASSINFOpdf
\else
\fi

\usepackage{comment}
\usepackage{ifthen}
\newboolean{showcomments}
\setboolean{showcomments}{true}
\newcommand{\ychen}[1]{\ifthenelse{\boolean{showcomments}}
        { \textcolor{red}{YC: #1}}}
\newcommand{\tongxin}[1]{\ifthenelse{\boolean{showcomments}}
        { \textcolor{blue}{(#1)}}{}}

\usepackage{amsmath}
\allowdisplaybreaks[4]

\begin{document}

%
\title{Hierarchical Game for Coupled Power System with Energy Sharing and Transportation System}
%
%
%

\author{Dongxiang Yan,
Tongxin Li,
Changhong Zhao,~\IEEEmembership{Senior Member,~IEEE},
Han Wang,
Yue Chen,~\IEEEmembership{Member,~IEEE}
\thanks{D. Yan and Y. Chen are with the Department of Mechanical and Automation Engineering, the Chinese University of Hong Kong, Hong Kong SAR. (email: dongxiangyan@cuhk.edu.hk, yuechen@mae.cuhk.edu.hk)}
\thanks{T. Li is with the School of Data Science, The Chinese University of
Hong Kong (Shenzhen), Shenzhen China. (email: litongxin@cuhk.edu.cn)}
\thanks{C. Zhao is with the Department of Information Engineering, the Chinese University of Hong Kong, Hong Kong SAR. (email: chzhao@ie.cuhk.edu.hk)}
\thanks{H. Wang is with the School of Electronic Information and Electrical Engineering, Shanghai Jiao Tong University, Shanghai 200240, China. (email: wanghan9894@sjtu.edu.cn)}
}
\markboth{Journal of \LaTeX\ Class Files,~Vol.~XX, No.~X, Feb.~2023}%
{Shell \MakeLowercase{\textit{et al.}}: Bare Demo of IEEEtran.cls for IEEE Journals}
%



\maketitle

\begin{abstract}
The wide deployment of distributed renewable energy sources and electric vehicles can help mitigate climate crisis. This necessitates new business models in the power sector to hedge against uncertainties while imposing a strong coupling between the connected power and transportation networks. To address these challenges, this paper first proposes an energy sharing mechanism considering AC power network constraints to encourage local energy exchange in the power system. Under the proposed mechanism, all prosumers play a generalized Nash game. We prove that the energy sharing equilibrium exists and is socially optimal. 
Furthermore, a hierarchical game is built to characterize the interactions both inside and between the power and transportation systems. Externally, the two systems are engaged in a generalized Nash game because traffic flows serve as electric demands as a result of charging behaviors, and each driver pays the energy sharing price for charging. The hierarchical game is then converted into a mixed-integer linear program (MILP) with the help of optimality conditions and linearization techniques. Numerical experiments validate the theoretical results and show the mutual impact between the two systems.
\end{abstract}


%
\IEEEpeerreviewmaketitle
\section*{Nomenclature}
\addcontentsline{toc}{section}{Nomenclature}
\subsection{Acronyms}
\begin{IEEEdescription}[\IEEEusemathlabelsep\IEEEsetlabelwidth{ssssssssss}]
\item [EV] {Electric vehicle.}
\item [GV] {Gasoline vehicle.}
\item [CS] {Charging station.}
\item [OD] {Origin-destination.}
\item [PV] {Photovoltaic.}
\item [TS] {Transportation system.}
\item [PS] {Power system.}
\item [DER] {Distributed energy resource.}
\item [MILP] {Mixed-integer linear program.}
\end{IEEEdescription}

\subsection{Indices and Sets}
\begin{IEEEdescription}[\IEEEusemathlabelsep\IEEEsetlabelwidth{ssssssssss}]
\item[$\mathcal{E}_N$] Set of nodes in the power system.
\item[$\mathcal{E}_L$] Set of lines in the power system.
\item[$\mathcal{C}$] Set of CSs.
\item[$C(i)$] Set of CSs powered by prosumer $i \in \mathcal{E}_N$.
\item[$\mathcal{T}_N$] Set of nodes in the transportation system.
\item[$\mathcal{T}_A$] Set of links in the transportation system.
\item[$\mathcal{T}_R$] Set of origin nodes.
\item[$\mathcal{T}_S$] Set of destination nodes.
\item[$\mathcal{K}^{rs}_{g}$] Set of GV paths connecting OD pair $rs$.
\item[$\mathcal{K}^{rs}_{e}$] Set of EV paths connecting OD pair $rs$.
\item[$T_A^R$] Set of regular links.
\item[$T_A^C$] Set of charging links.
\item[$T_A^B$] Set of bypass links.
\item[$S$] Set of partitions.
\end{IEEEdescription}
       
\subsection{Parameters}
\begin{IEEEdescription}[\IEEEusemathlabelsep\IEEEsetlabelwidth{ssssssssss}]
\item[$p_i$] Renewable generator output at prosumer $i \in \mathcal{E}_N$.
\item[$d_i^f$] Fixed power demand at prosumer $i \in \mathcal{E}_N$.
\item[$\underline{D}_i/\overline{D}_i$] Lower/Upper bound of elastic demand at prosumer $i$.
\item[$m_a$] Market sensitivity.
\item[$r_{ni},x_{ni}$] Resistance and reactance of line $(n,i)$.
\item[$l_{ni}$] Squared current magnitude of line $(n,i)$.
\item[$v_{n}$] Squared voltage magnitude of node $n$.
\item[$\underline{P_i}/\overline{P_i}$] Lower/Upper bound of $P_i$.
\item[$\underline{Q_i}/\overline{Q_i}$] Lower/Upper bound of $Q_i$.
\item[$\underline{v_i}/\overline{v_i}$] Lower/Upper bound of $v_i$.
\item[$\overline{\ell_{ni}}$] Upper bound of $\ell_{ni}$.
\item[$q^{rs}_g/q^{rs}_e$] Total traffic flow between $r \in \mathcal{T}_R$ and $s \in \mathcal{T}_S$ for GV/EV.
\item[$\delta_{akg}^{rs}/\delta_{ake}^{rs}$] Indicator variable for GV/EV. 
\item[$t_a^0$] Free travel time of link $a$.
\item[$h_a$] Free travel capacity of link $a$.
\item[$t_{a,\mu}^c$] Average service time at CS $c$ of link $a$.
\item[$t_{a,w}^c$] Maximum waiting time at CS $c$ of link $a$.
\item[$h_{a}^c$] Maximum allowable EV flow at CS $c$ of link $a$.
\item[$\gamma^c_a$] Electricity price of charging link $a$ with CS $c$.
\item[$E_b$] EV battery charging demand.
\item[$\omega$] A constant converting time to monetary value.
\item[$Z$] Integer parameter for linear approximation.
\item[$c_m,A_1,A_2$] Constant coefficient matrices in compact form.
\item[$D_{TS},B_2$] Constant coefficient matrices in compact form.
\item[$\lambda_{v,s}^{min}/\lambda_{v,s}^{max}$] Lower/Upper bounds of $\lambda_{v,s}$ for partition $s$.
\item[$D_{TS,s}^{min}/D_{TS,s}^{max}$] Lower/Upper bounds of $D_{TS,s}$.
\item[$|S|$] Partition number of McCormick envelope.
\end{IEEEdescription}

\subsection{Decision Variables}
\begin{IEEEdescription}[\IEEEusemathlabelsep\IEEEsetlabelwidth{ssssssssss}]
\item[$d_i$] Elastic demand at node $i \in \mathcal{E}_N$.
\item[$D_i^{T}$] Total charging demand of links $a \in C(i)$.
\item[$q_i$] Sharing quantity of node $i \in \mathcal{E}_N$.
\item[$\lambda_i$] Sharing price for node $i \in \mathcal{E}_N$.
\item[$b_i$] Bid of prosumer $i \in \mathcal{E}_N$ in the sharing market.
\item[$P_{ni}/Q_{ni}$] Active/Reactive power flow of line $(n,i)$.
\item[$U_i$] Utility of prosumer $i \in \mathcal{E}_N$.
\item[$f_{kg}^{rs}/f_{ke}^{rs}$] GV/EV traffic flow of a path $k \in \mathcal{K}^{rs}_{g}/\mathcal{K}^{rs}_{e}$.
\item[$x_a$] Aggregate traffic flow of link $a\in \mathcal{T}_A$.
\item[$x_{ag}/x_{ae}$] Aggregate GV/EV traffic flow of link $a\in \mathcal{T}_A$.
\item[$t_a$] Travel time on link $a \in \mathcal{T}_A$.
\item[$c_{kg}^{rs}$/$c_{ke}^{rs}$] Total cost of a path $k \in \mathcal{K}^{rs}_g/\mathcal{K}^{rs}_e$ for GV/EV.
\item[$u^{rs}_g/u^{rs}_e$] Minimal cost from $r \in \mathcal{T}_R$ to $s \in \mathcal{T}_S$ for GV/EV.
\item[$\xi^z,\eta^z$] Auxiliary variables.
\item[$y_{v}$] Vector variable in compact form.
\item[$\lambda_{v},\mu_{v}$] Dual variable vectors in compact form.
\item[$y_s$] Binary variable for partition $s\in S$.
\item[$\lambda_{v,s}$] Dual variable $\lambda_{v}$ for partition $s\in S$.
\item[$D_{TS,s}$] Variable $D_{TS}$ for partition $s\in S$.
\item[$\sigma$] Auxiliary variable to replace bilinear term.
\end{IEEEdescription}

\section{Introduction}

\IEEEPARstart{T}{he} electrification of transportation systems together with massive deployment of clean sources such as renewable energy has been regarded as an important means of carbon emission reduction \cite{kong2023spatial}. 
According to the Deloitte report, electric vehicle (EV) developed rapidly in the past decade and is expected to comprise 48\%, 27\%, and 42\% of light-duty vehicle market shares in China, the U.S., and Europe, respectively, by 2030 \cite{hamilton2020electric}. Meanwhile, over 81,000 distributed wind turbines had been installed in the U.S. during 2003-2017 with a total capacity of 1,076 MW \cite{orrell20162015}. The residential solar photovoltaic (PV) panels also grew dramatically from 3,700 MW in 2004 to 150,000 MW in 2014 \cite{agnew2015effect}. The impact of these trends is twofold: Firstly, the proliferation of distributed energy resources (DERs) provides conventional consumers the ability to produce, turning them into proactive prosumers. The power system is changing from an operator-centric model to a prosumer-centric model. Secondly, the prosperity of EVs tightens the coupling of the power and transportation systems. Therefore, it is necessary to analyze the interaction between coupled power and transportation systems while considering the electric grid transformation.

For the first issue, instead of being dispatched centrally, prosumers prefer actively participate in energy management by efficiently configuring their production and consumption \cite{parag2016electricity}. Many recent studies turned to peer-to-peer (P2P) energy sharing (trading) since it can unlock prosumers' flexibility to the maximum and provide enough incentive to get rid of the dependence on financial support \cite{tushar2020peer}. Typical energy sharing mechanisms can be divided into optimization-based ones, cooperative game-based ones, and noncooperative game-based ones. Optimization-based mechanisms begin with a centralized problem and implement energy sharing with the help of distributed optimization algorithms, such as alternating direction method of multipliers \cite{lyu2021fully}. To have a clearer economic interpretation, game theory is applied.
Cooperative game-based mechanisms provide proper allocation rules so that all prosumers will spontaneously act towards the social optimum. Despite some well-known methods such as the Shapley value \cite{long2019game}, nucleolus \cite{han2018incentivizing}, Nash bargaining \cite{cui2020community}, designing the allocation rule is challenging due to asymmetric information and privacy concerns. For non-cooperative game-based mechanisms, two typical models are Stackelberg games \cite{liu2017energy,liu2017online,xu2020data} and generalized Nash games \cite{chen2020approaching,le2020peer,chen2023energy}. Specifically, in the Stackelberg game \cite{liu2017energy}, the operator moves first to decide on the electricity price and then prosumers choose their strategies. Reference \cite{liu2017online} further considered prosumers' supply and demand ratios in the pricing decision of the operator \cite{liu2017online}. A data-driven method was utilized to find a nearly optimal pricing strategy \cite{xu2020data}. However, under the Stackelberg game setting, the flexibility of prosumers is limited to some extent since they are followers whose action sets are influenced by the operator (leader). In contrast, generalized Nash games allow prosumers to participate actively as leaders, which is the focus of this paper. A generalized demand function based mechanism was proposed with practical bidding algorithms \cite{chen2020approaching}. Network constraints were further considered \cite{le2020peer,chen2023energy}. 
{\color{black} A bilevel model was proposed to model the interaction between the grid operator and P2P energy trading market \cite{yang2023optimal}.}
However, they adopted simple DC power flow models, which are not accurate enough for the power distribution systems.
{\color{black} The AC network model was further integrated with a bilateral energy trading mechanism \cite{li2018distributed} and direct energy trading framework \cite{kim2019direct}.
Nevertheless, the above studies concentrated on analyzing the behavior of power systems only, whereas the interaction of multi-infrastructures implementing energy sharing remains unexplored.} 

For the second issue, a large number of EV charging stations (CSs) have been constructed in recent years to meet the growing EV charging needs. On the one hand, the routing of EV traffic flow for CSs will impact the road congestion of transportation systems \cite{zhang2020power}. On the other hand, the high charging demand from CSs can pose a threat to the distribution network operation \cite{tang2016model}.
Therefore, it is important to analyze the impact of EVs’ traveling and charging behavior on the two systems' operation \cite{zhang2018joint}. 
References \cite{wei2016optimal} and \cite{lv2019optimal} studied the optimal traffic-power flow using a mixed integer linear program (MILP). The EV scheduling strategies \cite{sun2018ev}, the urban transportation network expansion planning \cite{wei2017expansion}, 
and the resilience enhancement \cite{wang2018resilience} of coupled transportation and power systems were investigated. The above works assumed that both systems are under the control of the same operator. However, in reality, the two systems belong to different stakeholders and may have conflicting interests. A bi-level model was proposed to coordinate the two systems \cite{liu2023bilevel}. 
{\color{black} Reference \cite{yuan2023low} proposed a bi-level EV charging coordination approach to mitigate the global carbon emission of the coupled power and transportation systems.
A tri-level model was proposed to model the interaction among charging service providers, the transportation system and the power system \cite{li2022strategic}.
A pricing game was formulated for charging stations in the coupled power-transportation environment and solved by a deep reinforcement learning based method \cite{ye2023identifying}.
}
Reference \cite{wei2017network} considered the decision-making of each system as an independent stakeholder. Iterative methods were proposed to search for the equilibrium, which, however, have no convergence guarantee. Besides, their power systems adopted a centralized operation model without taking into account the promising trend of energy sharing in the prosumer era.
\begin{table}[htbp]\label{tab:summary}%
  \centering
  \color{black}
  \caption{Comparison of this paper with other related literature}
    \begin{tabular}{ccccc}
    \hline
    References  & \multicolumn{1}{c}{Energy} & \multicolumn{1}{c}{Power} & \multicolumn{1}{c}{Transportation} & Game \\
      & \multicolumn{1}{c}{sharing} & \multicolumn{1}{c}{system} & \multicolumn{1}{c}{system} &  \\
    \hline
    \cite{lyu2021fully}   & $\checkmark$     & x     & x     & x \\
    \cite{long2019game}--\cite{chen2020approaching} & $\checkmark$     & x     & x     & $\checkmark$ \\
    \cite{le2020peer}--\cite{kim2019direct} & $\checkmark$     & $\checkmark$     & x     & $\checkmark$ \\
    \cite{zhang2020power}--\cite{li2022strategic} & x     & $\checkmark$     & $\checkmark$     & x \\
    \cite{ye2023identifying,wei2017network} & x     & $\checkmark$     & $\checkmark$     & $\checkmark$ \\
    This paper & $\checkmark$ & $\checkmark$ & $\checkmark$ & $\checkmark$ \\
    \hline
    \end{tabular}  
\end{table}%

This paper studies the interaction of coupled power and transportation systems considering the strategic behavior within each system, i.e. the prosumers' strategic energy sharing behaviors in the power system and the EV drivers' strategic route selection behavior in the transportation system.
{\color{black}We summarize and compare the operation of the coupled power-transportation system proposed in this paper and other related literature in Table \ref{tab:summary}.}
The main contributions are twofold:

\emph{1) Hierarchical Game Model of Coupled Transportation and Power Systems.} We propose a hierarchical game model to characterize the complex interactions of coupled power and transportation systems. In the power system, an effective energy sharing mechanism is proposed under which all prosumers play a generalized Nash game. Different from the existing work, AC power network constraint is incorporated. In the transportation system, all drivers strategically select their routes to minimize their travel costs, which constitutes a Nash game. The charging load of EVs on a road with a charging station of the transportation system serves as the electric demand for a node in the power system, and the drivers pay for charging at the  prices derived from the energy sharing market. Hence, externally, there is a generalized Nash game between the two systems.

\emph{2) Method for Computing the Equilibrium.} To obtain the equilibrium of the proposed hierarchical game, we first provide two equivalent optimization models to compute the equilibrium in the power and transportation systems, respectively, in Propositions \ref{prop-1} and \ref{prop-2}. 
Then, to accelerate the computation, we linearize the nonlinear AC power network constraint via polyhedral approximation. Further, we replace the equivalent optimization model for energy sharing equilibrium in the power system with its \emph{primal-dual optimality} condition. 
The resulting nonlinear terms are linearized by piecewise McCormick envelope and SOS2 variables. Finally, the equilibrium of the hierarchical game can be solved by a MILP. Case studies show that the proposed method can greatly enhance the computational efficiency and avoid the disconvergence problem of the traditional methods. 



\section{Mathematical Formulation}
\begin{figure}[t]
\centering
\includegraphics[width=0.95\columnwidth]{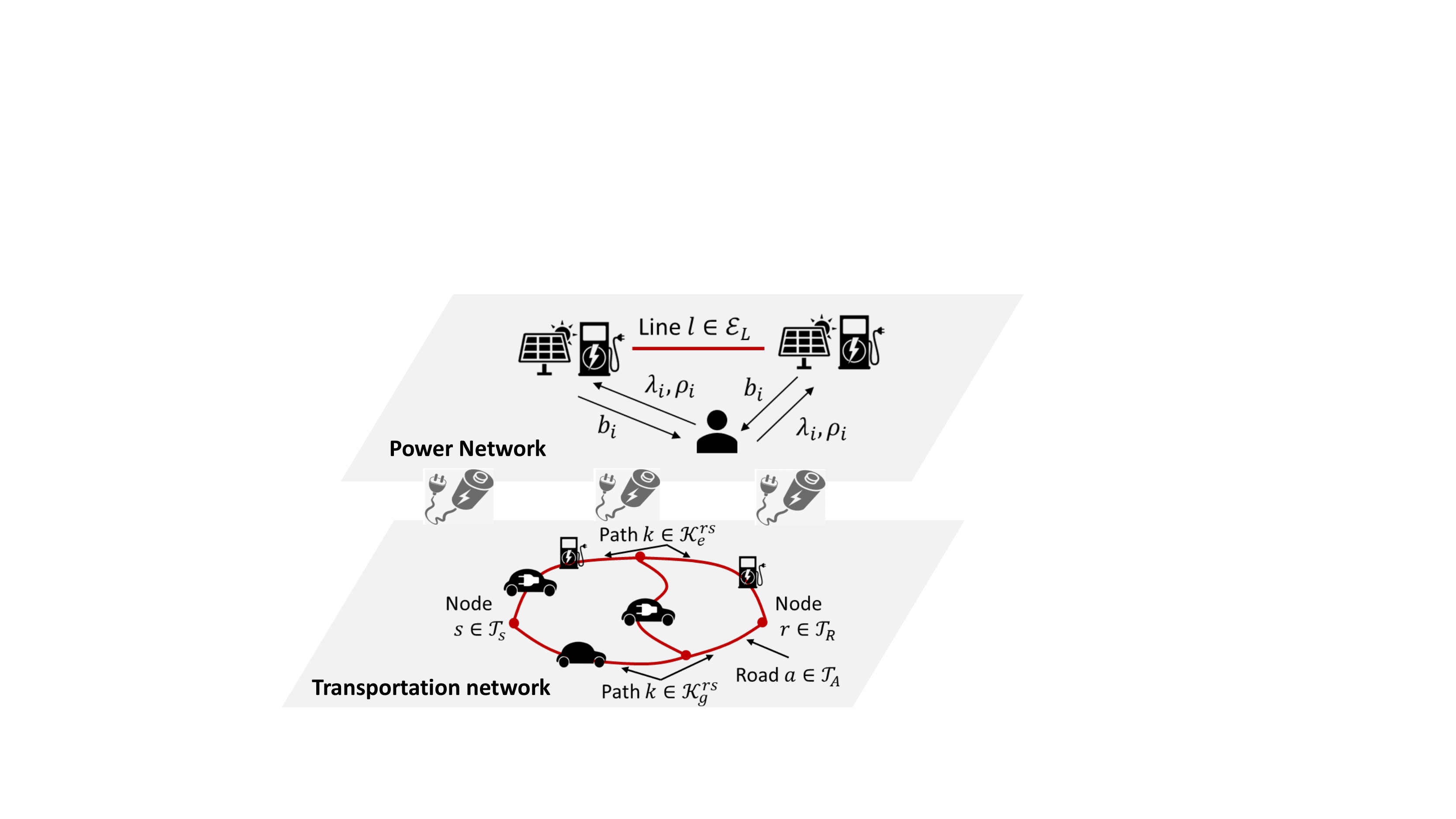}
\caption{Structure of the coupled power and transportation systems.}
\label{fig:structure}
\end{figure}

The structure of the coupled power and transportation systems is shown in Fig. \ref{fig:structure}. 
Each charging station (CS) in the transportation system is powered by a prosumer node in the power distribution system. When an EV goes to a CS, it receives a certain amount of energy. In the transportation system, we consider two kinds of vehicles: traditional gasoline vehicles (GV) and EVs. Each GV driver chooses the route that has the minimum travel time. Each EV driver chooses the route that can minimize its travel cost, which includes a cost proportional to the travel time and a cost related to the charging price.
In the power distribution system, each prosumer node is equipped with a distributed renewable generator to supply its demand and to provide electricity for a CS in the transportation system. The prosumer nodes can share energy with each other so that one with excessive power can sell electricity to another that lacks power. In general, the transportation system and the power system closely couple with each other. The traffic flow will influence the demand at the nodes of the power distribution system, which will further impact the energy sharing result. The energy sharing price affects the driving behavior of EV owners, which in turn influences the traffic flow in the transportation system. In the following, we will first introduce the game models inside the power and transportation systems, respectively; and then the hierarchical game model considering the interactions of the two systems.

{\color{black}
We introduce game theory into this problem because a game naturally exists among stakeholders in the power-transportation problem studied, i.e., among prosumers in the power grid and drivers in the transportation system, and between the power and the transportation system operators. Each stakeholder is rational and aims to maximize their own profit. As a result, conflicts of interests occur among stakeholders. It is thus well justified to analyze their strategic behaviors considering such interest conflicts, i.e., under a game setting. An alternative method, the centralized optimization method \cite{wei2016optimal} that assumes all stakeholders are controlled by a central operator, can achieve a lower total cost. However, the centralized method fails to account for the strategic behaviors and competing interests among stakeholders and has high computational burdens and privacy concerns. Therefore, a more practical method based on game theory is a must. There have been references studying the game equilibrium in power-transportation networks \cite{wei2017network,ye2023identifying}. For example, reference \cite{wei2017network} formulated a pricing game for charging service providers in a power-transportation network. Reference \cite{ye2023identifying} analyzed the equilibrium between the power and the transportation networks. Our paper extends and improves these studies.

While the existing studies focus on analyzing the game equilibrium, we move a step further -- design a market mechanism to coordinate the behaviors of strategic stakeholders so that their equilibrium could have a good economic efficiency. In particular, we propose an energy sharing market mechanism as follows. We prove that under the proposed mechanism, the energy sharing equilibrium can attain social optimum, as Proposition \ref{prop-1} states in Section \ref{sec-III}.

}

\subsection{Energy Sharing Equilibrium in Power System}
The power system can be modeled as a connected graph $\mathcal{G}_E=[\mathcal{E}_N,\mathcal{E}_L]$, where $\mathcal{E}_N$ and $\mathcal{E}_L$ represent the set of nodes and power lines, respectively. Each prosumer node $i$ has a renewable generator whose output is $p_i$, fixed demand $d_i^f$, and elastic demand $d_i \in [\underline{D}_i,\overline{D}_i]$. Other nodes only have fixed demand. For notation conciseness, for the nodes with fixed demand, we let their $p_i=0$ and $\underline{D}_i=\overline{D}_i=0$. Denote the set of CSs supplied by prosumer node $i$ as $C(i)$, which imposes total power demand $D_i^{T}$ by EVs. With the real-time renewable generator output $p_i$, the prosumer $i \in \mathcal{E}_N$ can adjust its elastic demand $d_i$ and share its excessive/deficit energy $q_i$ with other prosumers at a price $\lambda_i$ to maintain energy balancing. The utility of elastic demand can be represented by concave functions $U_i(d_i),\forall i$, so that $-U_i(d_i),\forall i$ are convex functions.
We propose an energy sharing mechanism as follows:

To participate in energy sharing, each prosumer $i \in \mathcal{E}_N$ submits a bid $b_i$ to the market operator, indicating its willingness to buy. The energy sharing market prices $\lambda_i,\forall i \in \mathcal{E}_N$ and the sharing quantities $q_i,\forall i \in \mathcal{E}_N$ will be determined according to the bids. Their relationship can be represented by a generalized demand function $q_i=-m_a\lambda_i+b_i$, where $m_a>0$ is the market sensitivity. It shows that given the same price $\lambda_i$, the prosumer will buy more if it has a higher willingness to buy (the higher $b_i$, the higher $q_i$); the prosumer will buy less when energy is more expensive (the higher $\lambda_i$, the less $q_i$). The market operator clears the market by solving:
\bsq
\label{eq:market}
\begin{align}
    \min_{\lambda_i,\forall
i \in \mathcal{E}_N}~ & \sum_{i \in \mathcal{E}_N} \lambda_i^2, \label{eq:market.1}\\
    \mbox{s.t.}
    ~ & \{-m_a\lambda_i+b_i,~\forall i \in \mathcal{E}_N\} \in \mathcal{F}_c, \label{eq:market.2}
\end{align}
\esq
where
\bsq
\label{eq:market.constr}
\begin{align}
    ~ & \mathcal{F}_c:=\{ q_i,~\forall i \in \mathcal{E}_N ~| \nonumber\\
    ~ & P_{ni}-q_i-r_{ni}\ell_{ni}-\sum\nolimits_{k:(i,k)}P_{ik}=0, \forall (n,i) \in \mathcal{E}_L,\label{eq:market.4} \\
    ~ & Q_{ni}+Q_{i}-x_{ni}\ell_{ni}-\sum\nolimits_{k:(i,k)}Q_{ik}=0,\forall (n,i) \in \mathcal{E}_L, \label{eq:maerket.5} \\
    ~ & v_{n}-2(r_{ni}P_{ni}+x_{ni}Q_{ni})+(r^2_{ni}+x^2_{ni})\ell_{ni}=v_{i},\forall (n,i) \in \mathcal{E}_L ,\label{eq:market.6} \\
    ~ & P_{ni}^2+Q_{ni}^2 \le \ell_{ni}v_{n},\forall (n,i) \in \mathcal{E}_L,\label{eq:market.7}\\
    ~ & \underline{P_i}\leq q_i \leq \overline{P_i}, \forall i \in \mathcal{E}_N ,\label{eq:market.8}\\
    ~ & \underline{Q_i}\leq Q_{i}\leq \overline{Q_i}, \forall i \in \mathcal{E}_N ,\label{eq:market.9}\\
    ~ & \underline{v_i}\leq v_{i}\leq \overline{v_i}, \forall i \in \mathcal{E}_N,\label{eq:market.10}\\
    ~ & 0\leq \ell_{ni}\leq \overline{\ell_{ni}},~\forall (n,i) \in \mathcal{E}_L\label{eq:market.11} ~\}.
\end{align}
\esq
where $Q_{i}$ is bus $i$'s reactive power injection, $P_{ni}/Q_{ni}$ is the active/reactive power flow of line $(n,i)$, $r_{ni}/x_{ni}$ is the resistance/reactance, and $\ell_{ni}/v_{n}$ is the squared current/voltage magnitude.
The objective function \eqref{eq:market.1} is designed to minimize the sum of square of the energy sharing prices. We will prove later in Proposition \ref{prop-1} that such a design can guarantee an equilibrium that is socially optimal. 
Constraints \eqref{eq:market.4}-\eqref{eq:market.7} describe the branch flow model. Second-order cone relaxation is applied, so the original equality constraint has been turned into the inequality constraint in \eqref{eq:market.7}. Constraints \eqref{eq:market.8}-\eqref{eq:market.11} include the physical bounds of prosumer power injection capacities, squared bus voltage magnitudes, and squared line current magnitudes. 

For each prosumer $i \in \mathcal{E}_N$, its net cost equals the energy sharing cost $\lambda_iq_i$ (when $q_i<0$, it means prosumer $i \in \mathcal{E}_N$ will sell energy to the market and receive revenue $-\lambda_iq_i$) minus its utility $U_i(d_i)$. It can decide on the optimal values of $d_i$ and $b_i$ to minimize its net cost: 
\bsq
\label{eq:eachnode}
\begin{align}
    \mathop{\min}_{b_i,d_i}~ & -U_i(d_i)+\lambda_i(-m_a\lambda_i+b_i), \label{eq:eachnode.1}\\
    \mbox{s.t.}~ & p_i-m_a\lambda_i+b_i=d_i+d_i^f+D_i^{T} ,\label{eq:eachnode.2} \\
    ~ & \underline{D}_i \le d_i \le \overline{D}_i. \label{eq:eachnode.3}
\end{align}
\esq
Constraint \eqref{eq:eachnode.2} shows that the renewable generation plus the energy it buys from the energy sharing market can satisfy its fixed and elastic demands. Constraint \eqref{eq:eachnode.3} is the demand adjustable range. 




The aforementioned energy sharing mechanism can protect the privacy of both prosumers and the operator to some extent, since the network constraints \eqref{eq:market.4}-\eqref{eq:market.11} are known only to the operator and the individual capacity constraint \eqref{eq:eachnode.3} is available only to the prosumer. If we take the market operator as a virtual player, then all prosumers and the operator play a generalized Nash game.
A major feature that distinguishes the generalized Nash game from a standard Nash game is that both the objective function and the constraints of one player are influenced by the strategies of the other players \cite{facchinei2010generalized}. Denote the objective function \eqref{eq:market.1} of the market operator as $\Gamma_M(\lambda)$ and its feasible set as $\mathcal{X}_M(b):=\{\lambda ~ | ~ \eqref{eq:market.2} ~ \mbox{is satisfied.}\}$. Denote the objective function \eqref{eq:eachnode.1} of each node as $\Gamma_i(\lambda_i,b_i,d_i)$ and its feasible set as $\mathcal{X}_i(\lambda_i):=\{(b_i,d_i) ~|~ \eqref{eq:eachnode.2}-\eqref{eq:eachnode.3}~\mbox{are satisfied.}\}$. 
{\color{black}The game consists of the following elements:
\begin{enumerate}[1)]
    \item The set of players, including prosumers $i\in \mathcal{E}_N=\{1,2,\ldots,I\}$ and the market operator $M$.
    \item Strategy sets: for prosumer $i$, it is $\mathcal{X}_i(\lambda_i)=\{(b_i,d_i)~|~\eqref{eq:eachnode.2}-\eqref{eq:eachnode.3}~\text{are satisfied.}\}$, and for market operator $M$, it is $\mathcal{X}_M(b)=\{\lambda~|~\eqref{eq:market.2}~\text{is satisfied.}\}$.
    \item Payoff functions: for prosumer $i$, it is $\Gamma_i(\lambda_i,b_i,d_i)$ and for the market operator, it is $\Gamma_M(\lambda)$.
\end{enumerate}
To be concise, denote the energy sharing game by $\mathcal{G}_1=\{(\mathcal{E}_N,M),(\mathcal{X}_i,\mathcal{X}_M ),(\Gamma_i,\Gamma_M)\}$.}

\begin{definition} (Energy Sharing Equilibrium)
A profile $(b^*,d^*,\lambda^*)$ is a \emph{generalized Nash equilibrium} (GNE) of the energy sharing game \eqref{eq:market} and \eqref{eq:eachnode} if and only if
\begin{align}\label{eq:ESQ}
    \lambda^*=~ & \mbox{argmin}_{\lambda \in \mathcal{X}_M(b^*)} \Gamma_M(\lambda) \nonumber\\
        (b^*,d^*)=~ & \mbox{argmin}_{b_i,d_i \in \mathcal{X}_i(\lambda_i^*)} \Gamma_i(\lambda^*_i,b_i,d_i),\forall i \in \mathcal{E}_N
\end{align}
\end{definition}

In general, generalized Nash games are difficult to analyze, and there is no theoretical guarantee for the existence and uniqueness of an equilibrium. Moreover, as part of the power demand $D_i^{T}$ comes from the transportation system, the interaction among the two systems constitutes another Nash game described later in Section II-C. This makes the problem even more sophisticated. In Section III, we will derive an equivalent optimization model in Proposition \ref{prop-1} for computing the generalized Nash equilibrium efficiently.

{\color{black}
\emph{Remark}: With the proliferation of distributed energy resources (DERs), passive consumers turn into proactive prosumers, calling for innovative market mechanisms to coordinate their flexibility and reduce their overall operational costs \cite{dimeas2014smart,parag2016electricity}. Energy sharing market has emerged as one of the potential solutions to this challenge \cite{wang2021distributed}. In an energy sharing market, prosumer makes decisions based on their own interests, thus giving rise to competition and forming a game. Usually, the equilibrium of such a game is not socially optimal because of the conflicts of interest. In this paper, we propose an energy sharing market mechanism with the market clearing rule in \eqref{eq:market}. In particular, \eqref{eq:market.1} is designed as minimizing the sum of squares of prices. We prove that under the proposed mechanism, the energy sharing equilibrium can attain social optimum, as Proposition \ref{prop-1} in Section \ref{sec-III} states.

}

\subsection{Wardrop User Equilibrium in Transportation System}
\label{secII-B}
The transportation system can be modeled as a connected graph $\mathcal{G}_T = [\mathcal{T}_N, \mathcal{T}_A]$, where $\mathcal{T}_N$ and $\mathcal{T}_A$ represent the set of nodes and roads (links), respectively. Each road (link) is denoted by $a \in \mathcal{T}_A$ and its traffic flow by $x_a$. Each driver needs to travel from an origin node $r$ to a destination node $s$.
Each origin-destination (OD) pair $rs$ is connected by a set of paths. We consider two kinds of vehicles in the transportation system: GVs and EVs. The GV traffic flow of a path $k \in \mathcal{K}^{rs}_g$ is $f_{kg}^{rs}$ and the given total GV traffic flow from $r$ to $s$ is $q^{rs}_g=\sum_{k \in \mathcal{K}^{rs}_g} f_{kg}^{rs}$. Similarly, for EV traffic flow, $q^{rs}_e=\sum_{k \in \mathcal{K}^{rs}_e} f_{ke}^{rs}$. To characterize the relationship between $f_{kg}^{rs}, f_{ke}^{rs},\forall k,\forall (rs)$ and $x_a$, we introduce the indicator variable $\delta_{akg}^{rs}$ ($\delta_{ake}^{rs}$) for GVs (EVs). $\delta_{akg}^{rs}=1$ ($\delta_{ake}^{rs}=1$) if road $a \in \mathcal{T}_A$ is part of the path $k \in \mathcal{K}^{rs}_g$ ($k \in \mathcal{K}^{rs}_e$); otherwise, $\delta_{akg}^{rs}=0$ ($\delta_{ake}^{rs}=0$). Therefore,
\begin{align}
\label{eq:x-f}
    x_a=~ \underbrace{\sum_{rs} \sum_{k \in \mathcal{K}^{rs}_g} \delta_{akg}^{rs} f_{kg}^{rs}}_{x_{ag}} + \underbrace{\sum_{rs} \sum_{k \in \mathcal{K}^{rs}_e} \delta_{ake}^{rs} f_{ke}^{rs}}_{x_{ae}},~ \forall a \in \mathcal{T}_A
\end{align}
where $\sum_{rs}$ is used to abbreviate $\sum_r\sum_s$. The travel time $t_a$ on a link $a \in \mathcal{T}_A$ depends on its aggregate traffic flow $x_a$.

\begin{figure}[t]
    \centering
    \includegraphics[width=0.55\columnwidth]{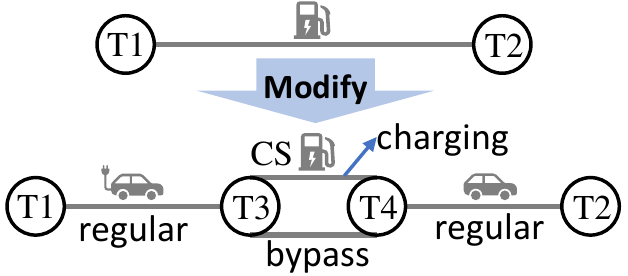}
    \caption{Illustration of three types of links.}\label{fig:3links}
\end{figure}

There are two types of links: the regular link without a CS $a\in T^R_A$ and the charging link with a CS. When an EV travels on the charging link, it can choose to get charged at the CS or bypass the CS. To facilitate the modeling and analysis, the original network is modified by adding a bypass link in parallel to the charging link, as shown in Fig. \ref{fig:3links}. The charging link is denoted by $a\in T^C_A$ while the bypass link by $a\in T^B_A$. $\mathcal{T}_A=T_A^R \cup T_A^C \cup T_A^B$.

For the regular link $a\in T^R_A$, we adopt the Bureau of Public Roads (BPR) function \eqref{eq:BPR} to describe the travel time, which increases with the traffic flow \cite{united1964traffic}.
\begin{align}
\label{eq:BPR}
    t_a(x_a) = t_a^0 [1+0.15(\frac{x_a}{h_a})^4]
\end{align}
where $t^0_a$ and $h_a$ are the free travel time and the capacity of link $a$.
For a charging link ($a\in T_A^C$), the time spent on it is
\begin{align}
\label{eq:t_cha}
    t_a(x_a) = t^{c}_{a,\mu}+t^{c}_{a,w}(\frac{x_a}{h_a^c})^3,~x_a\le h_a^c
\end{align}
where $t^{c}_{a,\mu}, t^{c}_{a,w}, h^c_a$ denote the average service time, the maximum waiting time, and the maximum allowable EV flow of the CS $c$ on link $a$. The travel time on a bypass link $a\in T_A^B$ is so short that is assumed to be zero, i.e., $t_a(x_a)=0,~\forall a \in T_A^B$.

For a GV owner, its travel cost $c^{rs}_{kg}$ on path $k\in K_{g}^{rs}$ between OD pair $rs$ is only influenced by the travel time:
\begin{align}
\label{eq:cost-gv}
    c_{kg}^{rs} = \sum\nolimits_{a \in \mathcal{T}_A} \omega t_a(x_a) \delta_{akg}^{rs}
\end{align}
where $\omega$ is the monetary cost of travel time.

For an EV owner, when deciding its route, it cares about two factors: the travel time and the charging price. 
Denote the charging price of the charging link $a$ with CS $c$ by $\gamma^c_a$, and we have $\gamma^c_a=\lambda_i$ if CS $c \in C(i)$. 
The EV travel cost of a path $k \in \mathcal{K}^{rs}_e$ is
\begin{align}
\label{eq:cost-ev}
    c_{ke}^{rs} = \sum \nolimits_{a \in \mathcal{T}_A} \omega t_a(x_a) \delta_{ake}^{rs}+ \sum \nolimits_{a \in T_A^C} \gamma^c_a E_b \delta_{ake}^{rs}
\end{align}
where $E_b$ represents an EV's charging demand, which is a constant, e.g., 20kWh.

Given the total traffic flows $q_{rs}$ of all OD pairs $rs$, each driver will choose the route that can minimize its own travel cost. Meanwhile, the decisions of all drivers will affect the aggregate traffic flow $x_a,\forall a \in \mathcal{T}_A$ and thus the travel time $t_a(x_a)$. 
An equilibrium is reached when no driver can reduce its travel cost by changing its route unilaterally.
\begin{definition} (Wardrop User Equilibrium \cite{wardrop1952road}) A flow $f^*$ is called a \emph{Wardrop user equilibrium} if and only if for each OD pair, all paths used by GVs (EVs) have the same travel cost that is no greater than the cost of any unused paths, i.e.
\begin{align}
    0 \le f_{kg}^{rs*} \perp (c_{kg}^{rs*}-u^{rs*}_g) \ge 0, \forall k \in \mathcal{K}^{rs}_g,\forall rs \label{eq:wardop-con-gv}\\
    0 \le f_{ke}^{rs*} \perp (c_{ke}^{rs*}-u^{rs*}_e) \ge 0, \forall k \in \mathcal{K}^{rs}_e,\forall rs \label{eq:wardop-con-ev}
\end{align}
\end{definition}

Take \eqref{eq:wardop-con-gv} as an example. For any used path $k \in \mathcal{K}^{rs}_g$, we have $f_{kg}^{rs*}>0$ and that $c_{kg}^{rs*}$ equals $u^{rs*}_{g}$. For any unused path, we have $f_{kg}^{rs*}=0$ and that the travel cost $c_{kg}^{rs*}$ is larger than or equal to $u^{rs*}_{g}$. The value of $u^{rs*}_g$ represents the minimum travel cost that is the same across all used paths from $r$ to $s$. As one driver's driving behavior may influence other drivers' travel time and travel cost, all drivers constitute a Nash game.
{\color{black}The game consists of the following elements:
\begin{enumerate}[1)]
    \item The set of players, including EV owners $\mathcal{S}_e=\{1,2,\ldots,I_e\}$ and GV owners $\mathcal{S}_g=\{1,2,\ldots,I_g\}$.
    \item Strategy sets: for EV owner $i\in\mathcal{S}_e$ driving from $r$ to $s$, it is $\mathcal{Y}_e=\{ f_{ke}^{rs},\forall k \in \mathcal{K}_{e}^{rs} | \sum_{k \in \mathcal{K}_e^{rs}} f_{ke}^{rs}=q_e^{rs}\}$; and for GV owner $i\in\mathcal{S}_g$ driving from $r$ to $s$, it is $\mathcal{Y}_g=\{ f_{kg}^{rs},\forall k \in \mathcal{K}_{g}^{rs} | \sum_{k \in \mathcal{K}_g^{rs}} f_{kg}^{rs}=q_g^{rs}\}$.
    \item Payoff functions: for EV owners $i\in\mathcal{S}_e$ from $r$ to $s$, it is $\Gamma_e(x_a(f_{ke}^{rs},\forall e, k, rs; f_{kg}^{rs},\forall g,k,rs),\forall a)$ given by \eqref{eq:cost-ev}; for GV $g$ from $r$ to $s$, it is $\Gamma_g(x_a(f_{ke}^{rs},\forall e, k, rs; f_{kg}^{rs},\forall g,k,rs),\forall a)$ given by \eqref{eq:cost-gv}.
\end{enumerate}
To be concise, denote by $\mathcal{G}_2=\{(\mathcal{S}_e,\mathcal{S}_g),(\mathcal{Y}_{e},\mathcal{Y}_{g}),(\Gamma_e,\Gamma_g)\}$ the strategic form of this Nash game.}

\subsection{Hierarchical Game of Coupled Systems}
In Sections II-A and II-B, we introduced the generalized Nash game in the power system and the Nash game in the transportation system, respectively. Externally, the power and transportation systems couple in two ways:

1) The charging demand in the transportation system is part of the electric load in the power system
\begin{align}
    D_i^{T}=\sum \nolimits_{c\in C(i)} x_a E_b,~a~\mbox{is the link where CS}~c~\mbox{locates}
\end{align}

2) The electricity price $\gamma^c_a$ on each CS $c$ equals the sharing price $\lambda_i$ at node $i \in \mathcal{E}_N$ if $c \in C(i)$. Since each prosumer node can choose to sell energy in the sharing market or serve the transportation system, to prevent arbitrage, the prices for both options should be equal.
\begin{figure}[t]
\centering
\includegraphics[width=0.8\columnwidth]{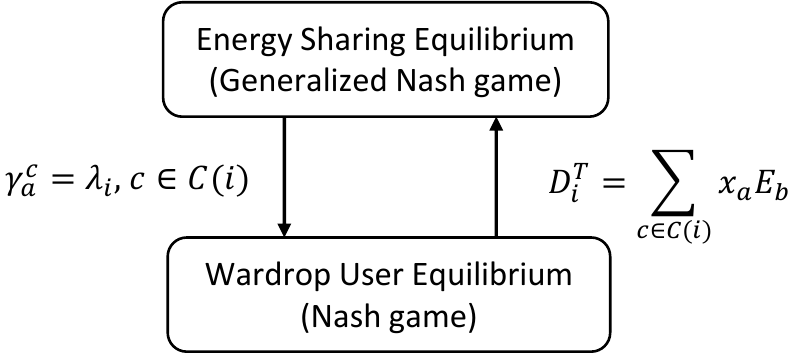}
\caption{Hierarchical game of transportation and power systems.}
\label{fig:game}
\end{figure}

The interactions inside and between the two systems are shown in Fig. \ref{fig:game}, which is a hierarchical game. To be specific, in the power system, there is a generalized Nash game among all prosumers participating in the energy sharing market. In the transportation system, all drivers strategically choose their route and play a Nash game. Moreover, the traffic flow in the transportation system will influence the demand in the power system, and the electricity price in the power system will impact the cost of EVs in the transportation system. Therefore, there is a \textcolor{black}{generalized Nash game} between two systems.
{\color{black}The game consists of the following elements:
\begin{enumerate}[1)]
    \item Players: the power system as a whole, denoted by $S_{ps}$ and the transportation system as a whole, denoted by $S_{ts}$.
    \item Strategy sets: for the power system, it is $\mathcal{Z}_{ps}(D_i^T,\forall i) = \{\lambda_i,\forall i~|~\lambda$ is the energy sharing price at the equilibrium \eqref{eq:ESQ}~\text{given}~$D_i^T,\forall i\}$; for the transportation system, it is  
    $\mathcal{Z}_{ts}(\lambda)=\{D_i^T,\forall i~|~D_i^T=\sum_{c \in \mathcal{C}_i}x_aE_b,~x_a$ is given by the Wardrop User Equilibrium with $\lambda$.
    \item Payoff functions: for power system, it is the sum of (3a) for all $i$ at the energy sharing equilibrium \eqref{eq:ESQ} given $D_i^T,\forall i$, denoted by   
    $\Gamma_{ps}(D_i^T,\forall i)$; for transportation system, it is the total travel cost of all EVs and GVs at the Wardrop User Equilibrium given $\lambda$, denoted by $\Gamma_{ts}$.
\end{enumerate}
To be concise, denote the game above by $\mathcal{G}_3=\{(S_{ps},S_{ts}),(\mathcal{Z}_{ps},\mathcal{Z}_{ts}),(\Gamma_{ps},\Gamma_{ts})\}$.}

\section{Solution Methodology}
\label{sec-III}
According to the analysis above, there is a hierarchical game between the coupled power and transportation systems. To the best of our knowledge, there is no existing method to analyze the equilibrium of such a hierarchical game due to its high complexity. To tackle this problem, in this section, we first provide two equivalent optimization models for computing the equilibrium of each system,  with theoretical proofs. Then we turn the external generalized Nash game between the two systems into a MILP based on optimality conditions and linearization techniques, to solve it efficiently.

\subsection{Optimization Models For Computing the Equilibrium}
We first provide two optimization models for computing the equilibrium inside the power and transportation systems, respectively, as stated in Propositions 1 and 2.

\begin{proposition}
\label{prop-1}
The generalized Nash equilibrium $(b^*,d^*,\lambda^*)$ of the energy sharing game \eqref{eq:market} and \eqref{eq:eachnode} can be obtained by
\bsq
\label{eq:share-eq}
\begin{align}
    \mathop{\min}_{d_i,\forall i \in \mathcal{E}_N} ~ & \sum \limits_{i \in \mathcal{E}_N} -U_i(d_i) \label{eq:share-eq.1}\\
    \mbox{s.t.} 
    ~ &  p_i+q_i=d_i+d_i^f+D_i^{T}, i \in \mathcal{E}_N ~:~\xi_i \label{eq:share-eq.2}\\
    ~ & \underline{D}_i \le d_i \le \overline{D}_i, \forall i \in \mathcal{E}_N~:~\psi_{i,d}^{\pm} \label{eq:share-eq.3}\\
   ~ & q \in \mathcal{F}_c
\end{align}
\esq
where $d_i^*,\forall i \in \mathcal{E}_N$ is unique with $\lambda_i^*=\xi^*_i$.
\end{proposition}

\begin{proposition}
\label{prop-2}
The traffic flow $f^*$ under Wardrop User Equilibrium can be obtained by
\bsq
\label{eq:traffic-eq}
\begin{align}
    \mathop{\min}_{x_a,f_{kg}^{rs},f_{ke}^{rs}}~ & \sum_{a\in \mathcal{T}_A} \int_0^{x_a} \omega t_a(\theta) d\theta + \sum_{a\in T^C_A} \gamma^c_a E_b x_a  \label{eq:traffic-eq.1}\\
    \mbox{s.t.} ~ & x_a=\sum_{rs} \sum_{k \in \mathcal{K}_g^{rs}} f_{kg}^{rs} \delta_{akg}^{rs} + \sum_{rs} \sum_{k \in \mathcal{K}_{e}^{rs}} f_{ke}^{rs} \delta_{ake}^{rs},\forall a \label{eq:traffic-eq.2}\\
    ~ & f_{kg}^{rs}\ge 0,\forall k \in \mathcal{K}^{rs}_g,\forall rs \label{eq:traffic-eq.3}\\
    ~ & f_{ke}^{rs}\ge 0,\forall k \in \mathcal{K}^{rs}_e,\forall rs \label{eq:traffic-eq.4}\\
    ~ & \sum \nolimits_{k \in \mathcal{K}^{rs}_g} f_{kg}^{rs}=q^{rs}_g,\forall rs ~:~u_g^{rs} \label{eq:traffic-eq.5}\\
    ~ & \sum \nolimits_{k \in \mathcal{K}^{rs}_e} f_{ke}^{rs}=q^{rs}_e,\forall rs ~:~u_e^{rs} \label{eq:traffic-eq.6}
\end{align}
\esq
where $x_a^*,\forall a \in \mathcal{T}_A$ is unique, and minimum travel cost $u^{rs*}_g$, $u^{rs*}_e$ equal the optimal dual variables of \eqref{eq:traffic-eq.5}, \eqref{eq:traffic-eq.6}, respectively.
\end{proposition}
The proofs of Propositions \ref{prop-1} and \ref{prop-2} are in Appendices \ref{apendix-A} and \ref{apendix-B}, 
respectively. They facilitate our further analysis. {\color{black}
It is worth noting that the property given by Proposition \ref{prop-1} is not trivial. In fact, at most of the time, the equilibrium of a generalized Nash game is not socially optimal; moreover, there is no guarantee on its existence and uniqueness \cite{facchinei2010generalized}. Even in a linear and convex setting, it is challenging to analyze a generalized Nash game due to the variability and interdependency of strategy sets, i.e., each player's strategy set depends on other players' strategies. In Proposition \ref{prop-1}, we prove that a unique energy sharing equilibrium exists and happens to be the optimal solution of the centralized problem \eqref{eq:share-eq}. This shows the effectiveness of the proposed energy sharing mechanism and facilitates further analysis of the market equilibrium.}

\subsection{Transformation and Linearization}
Two equivalent optimization models \eqref{eq:share-eq} and \eqref{eq:traffic-eq} are derived for computing the equilibrium in the power and transportation systems, respectively. As we can see from Fig. \ref{fig:game}, there is a Nash game between the two systems since the electricity price affects the travel cost in the transportation system and the traffic flow affects the demand in the power distribution system. In the following, we convert the optimization problems \eqref{eq:share-eq}-\eqref{eq:traffic-eq} into a set of mixed-integer linear constraints by optimality conditions and linearizations, based on which we can obtain the hierarchical game equilibrium of the coupled power and transportation systems. 

Regarding the energy sharing problem \eqref{eq:share-eq}, though the second-order cone constraint \eqref{eq:market.7} is convex, its nonlinearity still presents a challenge for efficient computation. To address this, polyhedral approximation is applied to turn \eqref{eq:market.7} into a set of linear constraints. 
{\color{black}Further, we replace the energy sharing problem with its primal-dual optimality condition, which is
different from the traditional method that is based on the
KKT condition. In contrast, the transportation problem \eqref{eq:traffic-eq} is
replaced by its KKT condition.
This is because the energy sharing problem comprises mainly of inequality constraints, for which the KKT condition would result in numerous complementary slackness constraints and require a lot of binary variables to turn it into a solvable mixed-integer linear program (MILP). The primal-dual optimality condition can avoid the high computational burden since it requires no binary variable. For the transportation problem with few inequality constraints, we still replace it with its KKT condition.}

\subsubsection{Primal-dual Optimality Condition of Problem \eqref{eq:share-eq}}
The polyhedral approximation technique in \cite{ben2001polyhedral} is employed to approximate the nonlinear constraint \eqref{eq:market.7} with linear constraints. First, constraint \eqref{eq:market.7} is rewritten as
\begin{equation}
     \sqrt{(2P_{ni})^2+(2Q_{ni})^2+(\ell_{ni}-v_n)^2}\le \ell_{ni}+v_n, \forall (n,i) \in \mathcal{E}_L, \label{eq:socp}
\end{equation}
which is further equivalently split into two second-order cones
\bsq
\begin{align}
    &\sqrt{(2P_{ni})^2+(2Q_{ni})^2}\le W_{ni}, \forall (n,i) \in \mathcal{E}_L,\\
    &\sqrt{W_{ni}^2+(\ell_{ni}-v_{n})^2}\le \ell_{ni}+v_n, \forall (n,i) \in \mathcal{E}_L.
\end{align}
\esq
Then, each constraint in the form of $\sqrt{x_1^2+x_2^2}\le x_3$ is approximated by a series of linear equalities and inequalities.
\begin{align}
~ &\left\{\begin{aligned}
        &\xi^0  \ge |x_1| \\
        &\eta^0 \ge |x_2|
        \end{aligned}\right.,~~ \left\{\begin{aligned}
        &\xi^Z \le x_3\\
        &\eta^Z \le \tan(\frac{\pi}{2^{Z+1}})\xi^Z
        \end{aligned}\right.\nonumber\\
~ &\left\{\begin{aligned}
        &\xi^z = \cos(\frac{\pi}{2^{z+1}})\xi^{z-1} + \sin(\frac{\pi}{2^{z+1}})\eta^{z-1} \\
        &\eta^z \ge \left|-\sin(\frac{\pi}{2^{z+1}})\xi^{z-1}+\cos(\frac{\pi}{2^{z+1}})\eta^{z-1}\right|
        \end{aligned},~z=1,...,Z\right. \label{eq:socplinset}
\end{align}
where $\xi^z,\eta^z,\forall z=0,1,2,...,Z,$ are auxiliary variables.
{\color{black}
The approximation error of the polyhedral approximation can be quantified by 
\begin{equation}
    \sqrt{(2P_{ni})^2+(2Q_{ni})^2+(\ell_{ni}-v_n)^2}\le [1+\epsilon(K)](\ell_{ni}+v_n),
\end{equation}
where
\begin{equation}
    \epsilon(K)=\frac{1}{\cos^2{\frac{\pi}{2^{K+1}}}}-1.
\end{equation}
The approximation error can be adjusted by parameter $Z$. A larger value of $Z$ leads to a smaller error of approximation at the cost of more auxiliary variables and a heavier computational burden.}
Finally, we use \eqref{eq:socplinset} to replace the nonlinear constraint \eqref{eq:market.7}. The convex objective function $-U_i(d_i),\forall i$ can also be linearized via a convex combination approach \cite{wu2011tighter}. Then, we get a LP-based energy sharing problem.

So far, the energy sharing game problem \eqref{eq:share-eq} can be cast as a compact form
\begin{align}
    \mathop{\min}_{y} ~  c_m^T y_{v},~
    \mbox{s.t.} ~ A_1 y_{v} = D_{TS} ~:\lambda_v, ~
     A_2 y_{v} \ge B_2 ~:\mu_v
\end{align}
where vector $y_{v}$ includes physical variables such as $q_i$, $Q_i$, $P_{ni}$, $Q_{ni}$, $v_i$, $\ell_{ni}$, $d_i$ and auxiliary variables $\xi^z,\eta^z$, and the matrices $c_m$, $A_1$, $A_2$, $D_{TS}$ and $B_2$ are constant coefficients. Vector $D_{TS}$ collects the transportation power demand $D_i^{T}$ of all prosumers. $\lambda_v$ is the dual variable associated with $A_1 y_v = D_{TS}$ and its j-th entry is the sharing price of prosumer $j$. All other equality and inequality constraints are included in $A_2 y_v \ge B_2$ with a dual variable vector $\mu_v$. Then, we can obtain the primal-dual condition
\bsq \label{eq:pd}
\begin{align}
    & A_1 y_v = D_{TS}, ~ A_2 y_v \ge B_2, \label{eq:pd.1}\\
    & A_1^T \lambda_v + A_2^T \mu_v = c_m, ~ \mu_v \ge 0, \label{eq:pd.2}\\
    & c_m^T y_v = \lambda_v^T D_{TS} + \mu_v^T B_2. \label{eq:pd.3}
\end{align}
\esq
The primal and dual constraints are represented by equations \eqref{eq:pd.1} and \eqref{eq:pd.2}, respectively. Due to the strong duality that always holds for LP, we have \eqref{eq:pd.3}, i.e., the optimal objective values of the primal and dual problems are equal.

\subsubsection{KKT Condition of Problem \eqref{eq:traffic-eq}}

The KKT condition is
\begin{gather}\label{eq:KKT2}
    \eqref{eq:x-f},\eqref{eq:cost-gv},\eqref{eq:cost-ev},\eqref{eq:wardop-con-gv},\eqref{eq:wardop-con-ev}
\end{gather}
For the above KKT condition \eqref{eq:KKT2}, nonlinearity lies in the complementary slackness condition and the travel time $t_a(x_a)$. The complementary slackness condition in the form of $0 \le x \perp y \ge 0$ can be linearized using the Big-M method.


\subsubsection{Linearization of the Bilinear Term and Travel Time}
When combing the conditions \eqref{eq:pd} and \eqref{eq:KKT2}, the presence of the bilinear product term $\lambda_v^T D_{TS}$ on the right-hand side of \eqref{eq:pd.3}, where $D_{TS}$ is a linear expression of $x_a$, makes the problem nonlinear again.
{\color{black}Such a term is approximated and convexified by McCormick envelope \cite{lv2022power}. The traditional version approximates the bilinear term by a set of linear functions over the domain between the lower and upper bounds of the variables. However, this may lead to a large approximation error when the domain is large. To address this issue, we adopt the piecewise McCormick envelope to tighten the error bounds. It partitions the variable domain into disjoint grids and obtains tighter linear bounds on each grid, thus improving the accuracy of approximation by engaging more auxiliary variables.}
The bilinear term is replaced by a new variable $\sigma$. Let $\lambda^{min}_v$ and $\lambda^{max}_v$ represent the lower and upper bounds of variable $\lambda_v$. $|S|$ is the partition number and $s$ is partition index. Binary variable $y_{s}=1$ means $\lambda_{v,s}$ belong to this disjunction; otherwise, $y_{s}=0$. $D_{TS}$ is also partitioned as $D_{TS,s}, \forall s\in S$.
\bsq \label{eq:Mccormick}
\begin{align}
    &\sigma \ge \sum\nolimits_{s=1}^{|S|}(D_{TS}^{min}\lambda_{v,s}+\lambda_{v,s}^{min}D_{TS,s}-D_{TS}^{min}\lambda_{v,s}^{min}y_{s}), \forall s \in S\\
    &\sigma \ge \sum\nolimits_{s=1}^{|S|}(D_{TS}^{max}\lambda_{v,s}+\lambda_{v,s}^{max}D_{TS,s}-D_{TS}^{max}\lambda_{v,s}^{max}y_{s}), \forall s \in S\\
    &\sigma \le \sum\nolimits_{s=1}^{|S|}(D_{TS}^{max}\lambda_{v,s}+\lambda_{v,s}^{min}D_{TS,s}-D_{TS}^{max}\lambda_{v,s}^{min}y_{s}), \forall s \in S\\
    &\sigma \le \sum\nolimits_{s=1}^{|S|}(D_{TS}^{min}\lambda_{v,s}+\lambda_{v,s}^{max}D_{TS,s}-D_{TS}^{min}\lambda_{v,s}^{max}y_{s}), \forall s \in S\\
    &\lambda_v = \sum\nolimits_{s=1}^{|S|}\lambda_{v,s},~D_{TS} = \sum\nolimits_{s=1}^{|S|}D_{TS,s},~ \sum\nolimits_{s=1}^{|S|}y_{s}=1, \forall s \in S\\
    &\lambda_{v,s}^{min} = \lambda_v^{min}+(\lambda_v^{max}-\lambda_v^{min})(s-1)/|S|, \forall s \in S\\
    &\lambda_{v,s}^{max} = \lambda_v^{min}+(\lambda_v^{max}-\lambda_v^{min})s/|S|, \forall s \in S\\
    &\lambda_{v,s}^{min}y_s \le \lambda_{v,s} \le \lambda_{v,s}^{max}y_s, \forall s \in S\\
    &D_{TS,s}^{min}y_s \le D_{TS,s} \le D_{TS,s}^{max}y_s, \forall s \in S\\
    &y_s\in\{0,1\}, \forall s \in S
\end{align}
\esq


After above transformation, the only nonlinear terms exist in travel time $t_a(x_a)$, i.e., \eqref{eq:BPR} and \eqref{eq:t_cha}. In this paper, we perform piecewise linear approximation by using SOS2 variables \cite{beale1976global}, which is a vector of variables with at most two adjacent elements being able to take nonzero values. The specific processes are omitted for brevity.
{\color{black} The potential error grows when the true nonlinear function deviates significantly from the applied linear segments. Generally, the approximation errors can be mitigated by increasing the number of linear segments and/or using adaptive segment placement.}
After the above transformation based on optimality conditions and the linearization techniques, the equilibrium of the hierarchical game in Fig. \ref{fig:game} can be calculated by solving a MILP.

{\color{black}
\subsection{Discussions on Practical Issues}
In the following, We further discuss some practical issues regarding the proposed model and method below:

\subsubsection{Performance guarantee and real-time issue}
We build a game model to characterize the interaction between the power and transportation systems. To compute the equilibrium of the hierarchical game, linearization techniques and optimality conditions are applied to convert the game into a MILP, which is then solved by Gurobi.
It is worth noting that ``real-time'' in this paper is a concept compared to the traditional ``day-ahead'' scheduling problem in the power system which covers 24 hours. The time interval of ``real-time'' in this paper refers to 1 hour, i.e., we study the problem with a single time period. This setting is widely adopted in references such as \cite{cui2021optimal} and \cite{wei2017network}.
In future work, we may take into account the temporal evolution of EV flows by using a multi-period power system model and a dynamic traffic assignment model.

\subsubsection{Joint ride and energy sharing}
Energy sharing occurs among prosumers of the power system, such as charging stations with renewable generation. Ride sharing occurs among vehicles in the transportation system. Generally, ride sharing can reduce the traffic flow and congestion, leading to lower travel time and costs. Meanwhile, the decreased EV traffic flow will reduce the charging demand and further affect the energy sharing strategies of prosumers as well as the (locational) electricity prices. In contrast, the electricity prices may affect the distribution of EV charging demands, and thus the traffic flows and the ride sharing patterns. Therefore, it is promising to take into account joint energy sharing and ride sharing in the future work.

\subsubsection{Utility function design}
The utility function represents the level of satisfaction of the user as a function of its total power consumption. According to reference \cite{samadi2012advanced},  the utility function $U_i(d_i)$ of prosumer $i$ should satisfy the following properties:

Property 1: Utility functions are nondecreasing. This implies that the marginal utility is nonnegative:
\begin{equation}
    \frac{\partial U_i}{\partial d_i}\ge 0.    
\end{equation}

Property 2: The marginal utility is a nonincreasing function of consumption, which implies the utility function is concave, or (when twice differentiable):
\begin{equation}
    \frac{\partial^2 U_i}{\partial d_i^2}\le 0.    
\end{equation}

Property 3: When the consumption $d_i=0$, 
\begin{equation}
U_i (0)=0.    
\end{equation}
A common simple example of $U_i (d_i)$ satisfying the conditions above is just a linear function \cite{chen2019energy}, which is used in this paper.

\subsubsection{Energy sharing equilibrium seeking}
The energy sharing equilibrium between prosumers and the market operator can be reached via a bidding algorithm in a distributed iterative manner. It includes two steps: operator update and prosumer update.
In the step of operator update, given the energy-sharing bids sent by prosumers, the market operator solves the problem \eqref{eq:market} to minimize the variation of the sharing prices across prosumers and clears the market.
In the step of prosumer update, each prosumer decides on its shared energy bid to minimize its net cost by solving the problem \eqref{eq:eachnode} using the up-to-date sharing price determined by the market operator.
The above procedures are repeated until convergence, which is illustrated in Algorithm \ref{algo:bidding} as follows. The effectiveness of the algorithm is also verified in the simulation of Section \ref{sec:IV-C}.

\begin{algorithm}[!htbp]
{\color{black}
\caption{{\color{black}Energy Sharing Equilibrium Seeking Algorithm}}\label{algo:bidding}
\begin{algorithmic}[1]
\STATE{Input the parameters of prosumer $i\in\mathcal{E}_N$}
\STATE{Set iteration index $k=1$, convergence tolerances $\delta>0$, sharing price $\lambda_i^k>0$.}
\REPEAT
\renewcommand{\algorithmicrequire}{ \textit{\textbf{~~~~~~prosumer update:}}}
\REQUIRE
\FOR{each prosumer $i\in\mathcal{E}_N$}
\STATE{solve the problem (3) to obtain the bid $b_i^k$ and scheduled demand $d_i^k$.}
\ENDFOR
\renewcommand{\algorithmicrequire}{ \textit{\textbf{~~~~~~operator update:}}}
\REQUIRE
\STATE{Market operator solves the problem (1) to obtain the sharing prices $\lambda_i$.}
\STATE{$k \leftarrow k+1$}
\UNTIL{$\|b^{k+1}-b^k\| \le \delta$.}
\end{algorithmic}}
\end{algorithm}
}

\section{Case Studies}
Numerical experiments are carried out to validate the proposed method and propositions.
All simulations are implemented in MATLAB, solved by Gurobi, and run on a laptop with an Intel Core i7 1.80 GHz CPU and 16 GB RAM.

\subsection{System Setup}
\begin{figure}[!htbp]
  \centering
\includegraphics[width=0.4\textwidth]{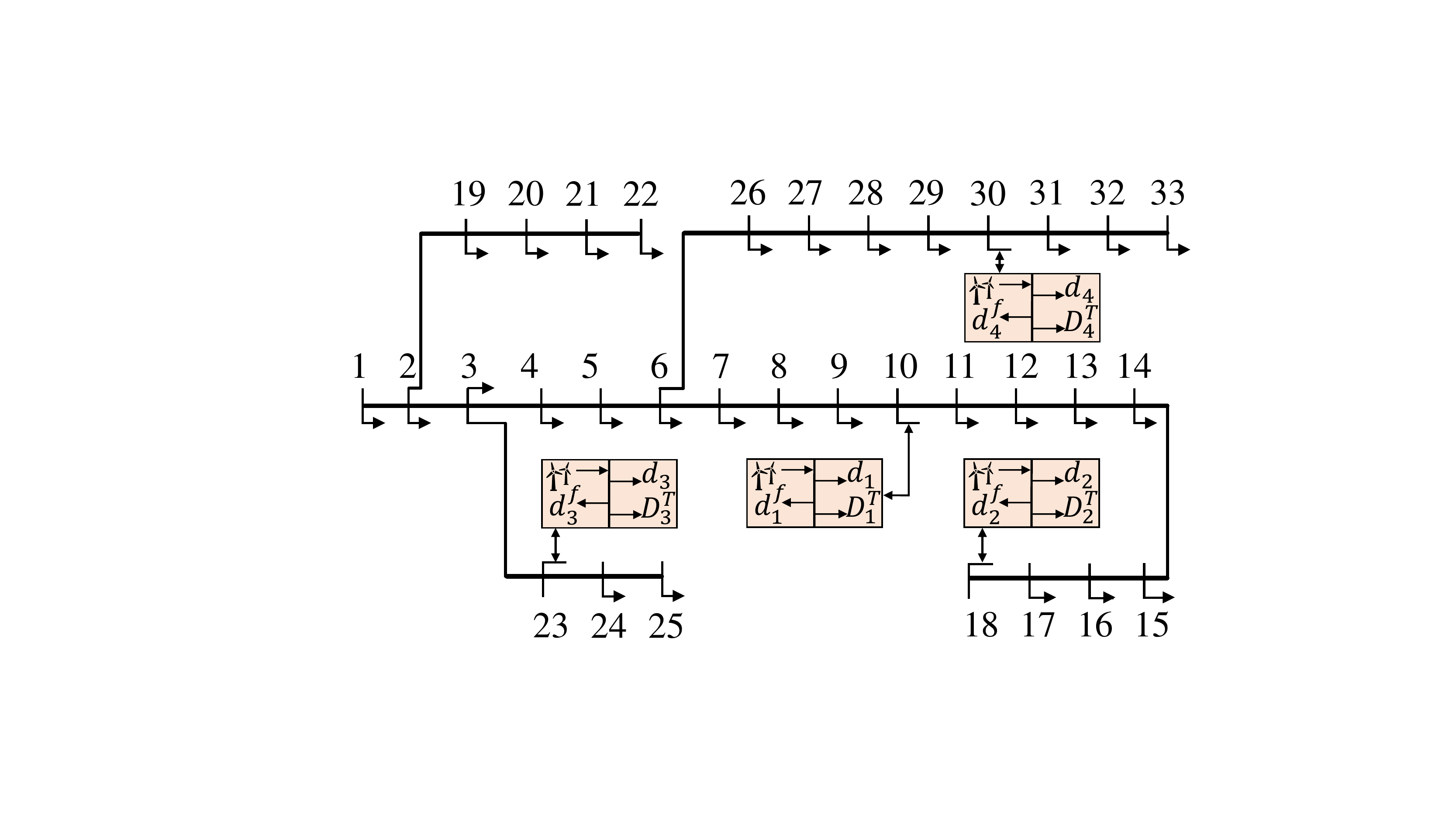}\\
  \caption{Modified IEEE 33-bus test system with 4 prosumers.}\label{fig:ieee33}
\end{figure}

\begin{figure}[!htbp]
  \centering
  \includegraphics[width=0.49\textwidth]{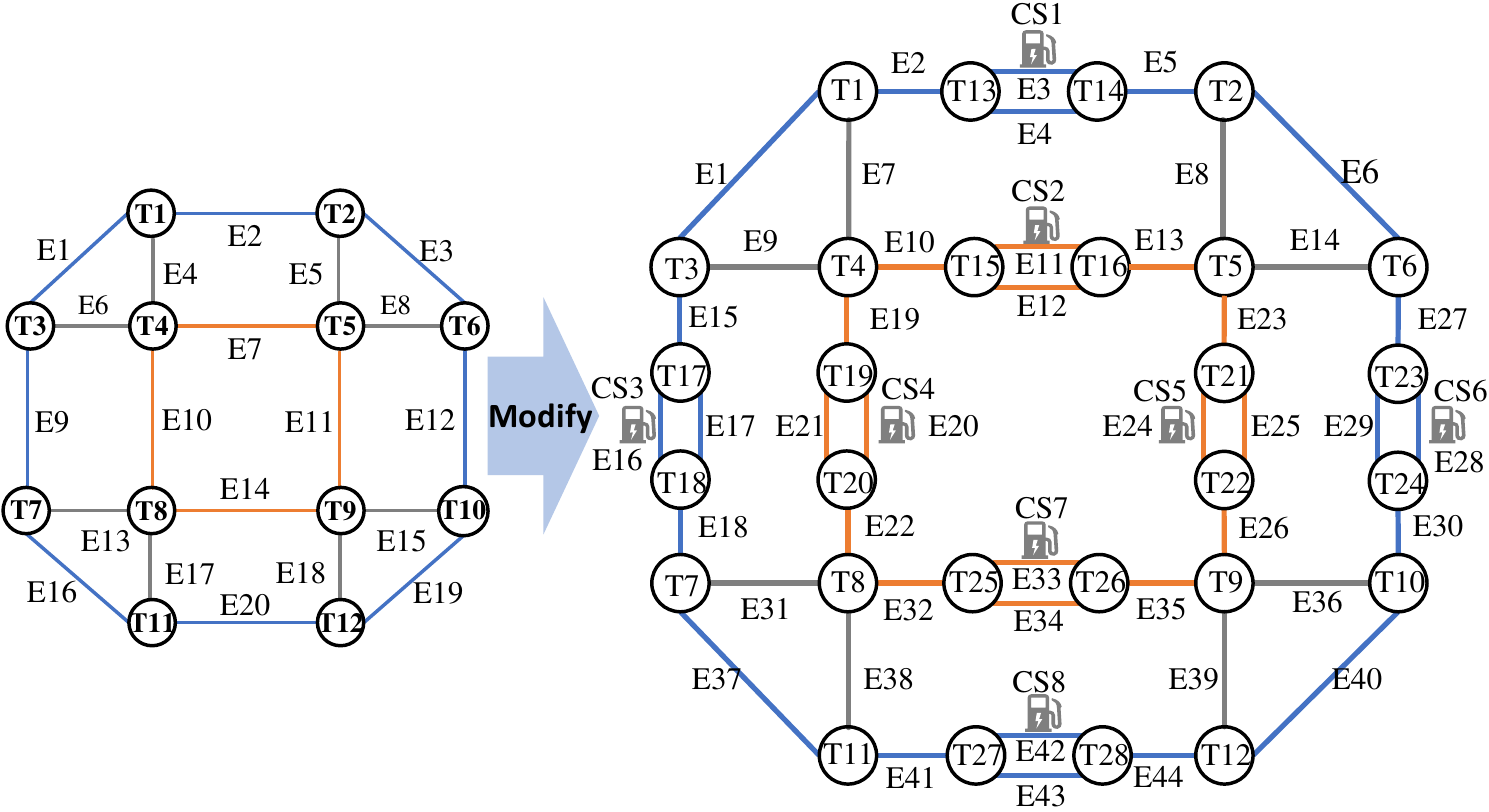}\\
  \caption{Left: Original transportation network with 12 nodes and 20 links. Right: Expanded transportation network with 28 nodes and 44 links.}\label{fig:TN}
\end{figure}

A modified IEEE 33-bus system is adopted as the power network as shown in Fig. \ref{fig:ieee33}. We first consider the case with four prosumers located at buses 10, 18, 23, and 30, respectively. Each prosumer possesses renewable generation ($p_1/p_2/p_3/p_4$=3/1/4/2MW), fixed load ($d_1/d_2/d_3/d_4$=0.1/0.2/0.15/0.1MW), adjustable load, and powers one or multiple charging stations (CSs). The upper and lower limits of voltage magnitudes for each bus are 1.06 p.u. and 0.94 p.u., respectively.
Fig. \ref{fig:TN} (left) shows the transportation network (TN) with 12 nodes and 20 links.
The parameters of each link can be found in \cite{lv2022power}. The OD pairs and trip rates are listed in TABLE \ref{tab:odinfo}. The base value of traffic flow is 100 vehicles/hour and we assume that EVs account for 5\% of the trip rate of each OD pair. As discussed in Section \ref{secII-B}, bypass links are added to facilitate the analysis.
As a result, it is expanded to a 28-node and 44-link network, as shown in Fig. \ref{fig:TN} (right).
Based on the expanded transportation network, we then build its node-link incidence matrix and enumerate the paths for GVs and EVs between OD pair $(r,s)$ using the method in \cite{wei2017network}.
Regarding the coupling between power and transportation networks, we assume CS1 and CS2 are powered by prosumer 1, CS3 and CS4 by prosumer 2, CS5 and CS6 by prosumer 3, and CS7 and CS8 by prosumer 4.
Each EV's charging demand is assumed to be $E_b=20$kWh and charging time is $t^c_{a,\mu}=20$min. The monetary cost of travel time is $\omega=10$\$/hr.

\begin{table}[htbp]
  \centering
  \caption{OD pairs and their trip rates (in p.u., 1p.u.=100 vehicles/h).}\label{tab:odinfo}%
    \begin{tabular}{cc|cc|cc|cc}
    \hline
    OD pair & $q_{rs}$ & OD pair & $q_{rs}$ & OD pair & $q_{rs}$ & OD pair & $q_{rs}$ \\
    \hline
    T1-T6  & 5   & T1-T10 & 15   & T1-T12  & 10  & T1-T11 & 5\\
    T3-T6  & 5   & T3-T10 & 15   & T3-T12  & 15  & T3-T11 & 5\\
    T4-T9  & 10  & T4-T10 & 10   & T4-T12  & 5   &        &\\
    \hline
    \end{tabular}%
\end{table}%

\subsection{Equilibrium Results}
The equilibrium of the external generalized Nash game between the power and transportation systems is characterized by the sharing/charging price $\lambda_i^*$ and traffic flow $x_a^*$. Moreover, they are the results of the two internal (generalized) Nash equilibria.
At the equilibrium, we compute the difference between the left-hand side and the right-hand side of \eqref{eq:socp} under $Z=6$. The errors associated with all lines fall within the range of $-0.85\times 10^{-3}$ to $2.18\times 10^{-3}$, indicating that the proposed polyhedral approximation can achieve desirable accuracy. The computation takes 120 seconds.

\begin{figure}[!htbp]
  \centering
  \includegraphics[width=0.4\textwidth]{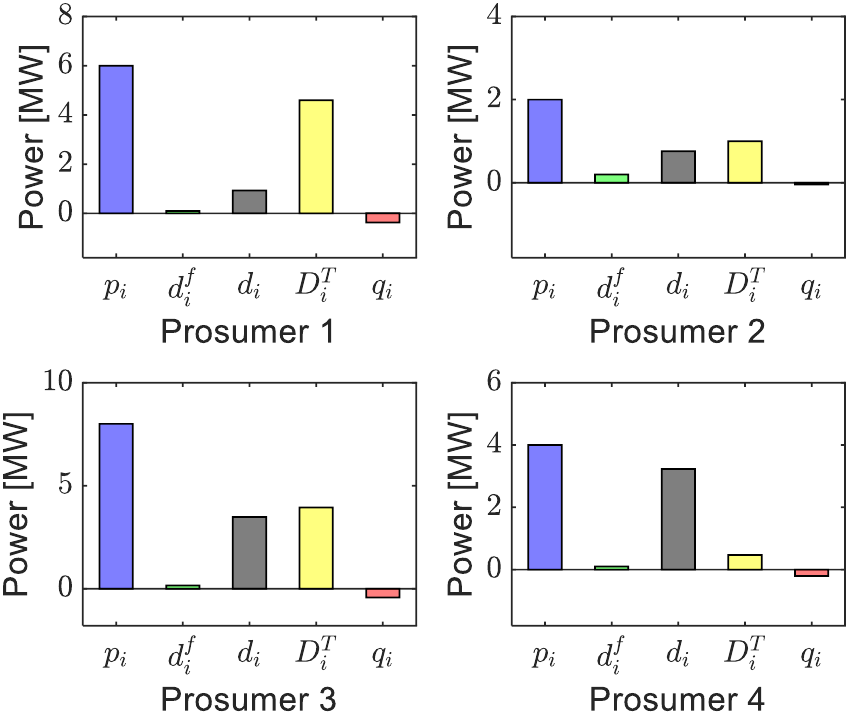}\\
  \caption{Power distribution within each prosumer.}\label{fig:4power}
\end{figure}

We first discuss the energy sharing equilibrium in the power network. At the equilibrium point, the power distribution within each prosumer is shown in Fig. \ref{fig:4power}. 
All sharing quantity $q_i <0,\forall i$, indicating that the four prosumers 1-4 are all sellers in the energy market. The sum of $|q_1|$ to $|q_4|$ corresponds to the injected power into the power network for serving the loads located at other buses and compensating for the power loss in the AC distribution network.
The energy sharing price of each prosumer is $\lambda_1=0.4336$\$/kWh, $\lambda_2=0.4402$\$/kWh, $\lambda_3=0.4342$\$/kWh, and $\lambda_4=0.4400$\$/kWh. 
The prices determined by energy sharing are also used as the EV charging prices in the transportation system. Owing to the difference in charging prices, it can be seen that prosumers with lower prices will serve a larger EV charging load and vice versa. For instance, prosumer 1 serves the largest EV charging load of 4.5975 MWh because it offers the lowest charging price of 0.4336 \$/kWh. 

\begin{figure}[!htbp]
  \centering
  \includegraphics[width=0.4\textwidth]{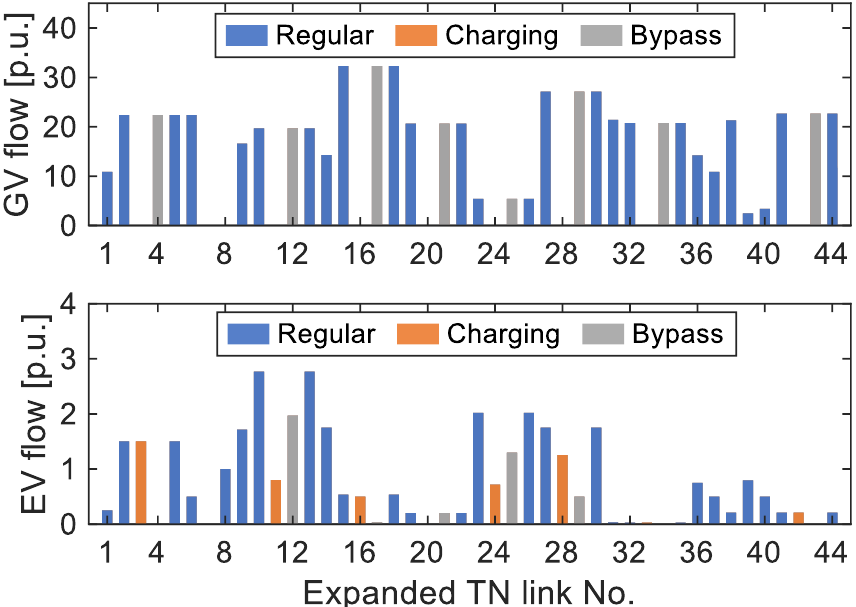}\\
  \caption{GV and EV traffic flow assignment in each expanded TN link.}\label{fig:gvevflow}
\end{figure}

As for the Nash game in the transportation system, the equilibrium is reached when \eqref{eq:traffic-eq} is solved. Fig. \ref{fig:gvevflow} depicts the traffic flows of GVs and EVs across each expanded TN link. As seen, GVs solely utilize regular and bypass links while the traffic flow of EVs is distributed across all three types of links.
To demonstrate the reached equilibrium, we take OD pair T1-T12 for example. Under the UE, for OD pair T1-T12, all EVs select the most cost-effective path (E2-E3-E5-E6-E27-E29-E30-E40), with a travel cost of \$19.45, including a travel time cost of \$10.78 and a charging cost of \$8.67. On the contrary, if an EV selects the path (E7-E10-E11-E13-E23-E25-E26-E39), the corresponding travel cost would be \$25.57. This cost comprises a travel time cost of \$16.9 and a charging cost of \$8.67. The two paths have the same charging cost because both CS1 and CS2 are powered by the same prosumer 1, but different travel cost. The second path is more expensive than the first one that is the optimal.
However, for GVs traveling between the OD pair T1-T12, there are two optimal paths. The first path is (E2-E4-E5-E6-E27-E29-E30-E40) and the second is (E1-E15-E17-E18-E37-E41-E43-E44). Both of them have the same travel cost of \$7.41. Notably, this cost is less than those associated with the non-selected paths, such as (E7-E10-E12-E13-E23-E25-E26-E39) that costs \$13.43. These results align with the definition of Wardrop UE.

{\color{black}
\subsection{Distributed Energy Sharing Mechanism Verification}\label{sec:IV-C}
Proposition \ref{prop-1} states that the GNE of energy sharing game \eqref{eq:market} and \eqref{eq:eachnode} can be obtained by solving the centralized problem \eqref{eq:share-eq}.
To verify Proposition \ref{prop-1}, we first fix the transportation power demand $D_i^{T}$ of each prosumer and solely solve the corresponding centralized problem \eqref{eq:share-eq}. Under this setting, we can obtain prosumers' elastic loads: $d_1=0.93$MW, $d_2=0.75$MW, $d_3=3.47$MW, $d_4=3.23$MW, and their sharing prices by retrieving the dual variables of \eqref{eq:share-eq.1} with Gurobi: $\lambda_1=0.41$\$/kWh, $\lambda_2=0.42$\$/kWh, $\lambda_3=0.43$\$/kWh, $\lambda_4=0.44$\$/kWh. The prices are slightly different from those mentioned previously because of the linearization and approximation errors.
Then, we let the market operator and prosumers play the generalized Nash game according to the distributed Algorithm \ref{algo:bidding}. The changes in energy sharing prices and elastic demands are illustrated in Fig. \ref{fig:gne}. As the iteration proceeds, we can see that both the energy sharing prices and prosumers' strategies of load adjustment amount gradually converge. At GNE, we have $d_1=0.91$MW, $d_2=0.77$MW, $d_3=3.49$MW, $d_4=3.22$MW, and $\lambda_1=0.41$\$/kWh, $\lambda_2=0.42$\$/kWh, $\lambda_3=0.43$\$/kWh, $\lambda_4=0.44$\$/kWh. They are almost the same as the centralized solution. Therefore, Proposition \ref{prop-1} and Algorithm \ref{algo:bidding} are verified.

\begin{figure}[!htbp]
  \centering
  \includegraphics[width=0.4\textwidth]{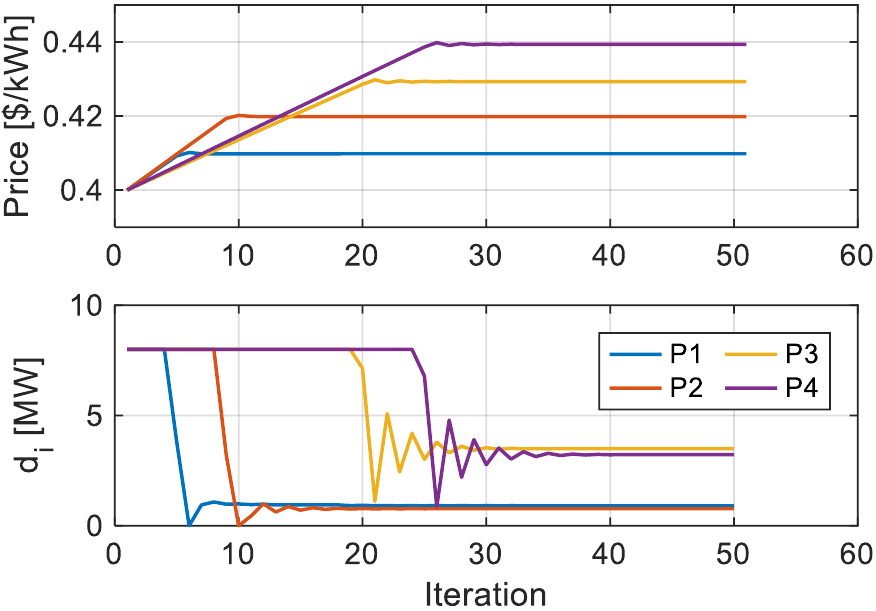}\\
  \caption{Energy sharing prices and optimal strategies of four prosumers. ``P'' is the abbreviation of prosumer.}\label{fig:gne}
\end{figure}

}

\subsection{Performance Comparison}\label{sec:comparison}
To demonstrate the advantage of the proposed approach, three benchmarks used in the literature are conducted and compared with the proposed method.
\begin{itemize}
  \item Benchmark 1 (B1): Best response decomposition method \cite{wei2017network}. Iteratively and alternatively, it  solves the energy sharing of the power network by fixing the charging demand as given by the traffic, and then fixes the energy-sharing prices to solve the traffic flow assignment. However, this method may not converge.
  \item Benchmark 2 (B2): Traditional KKT conditions method. Both energy sharing equilibrium and the Wardrop user equilibrium are replaced with their KKT conditions.
  \item Benchmark 3 (B3): Partial energy sharing. We assume some prosumers do not participate in energy sharing but operate in a self-sufficiency mode, i.e., $q_i=0$. Other settings are the same as the proposed method.
  \item Proposed method: All prosumers participate in energy sharing and we solve the hierarchical game via the derived MILP without iteration.
\end{itemize}

\begin{figure}[!htbp]
  \centering
  \includegraphics[width=0.4\textwidth]{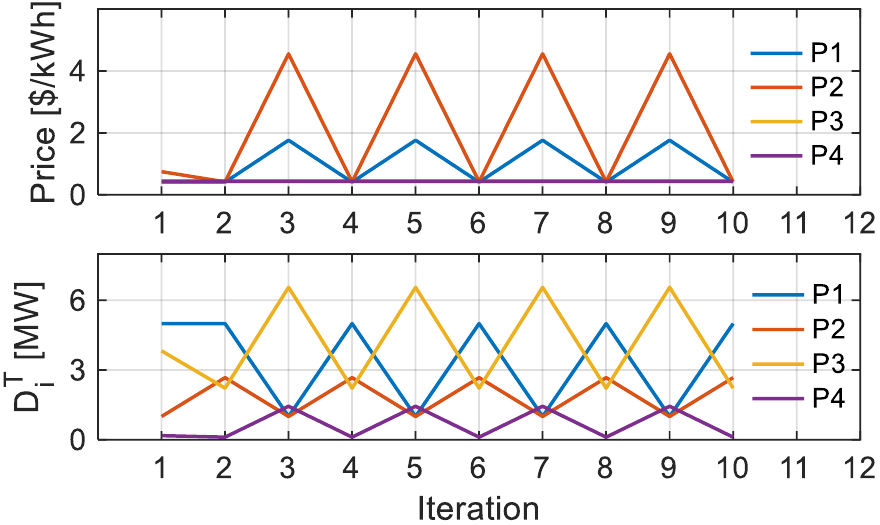}\\
  \caption{The oscillation of $D_i^{T}$ and $\lambda_i$ under B1.}\label{fig:B1}
\end{figure}

For B1, in our test, if we keep the power network constraints (e.g., bus voltage limits and line flow limits) the same as those used in the proposed method, B1 stops after 1 iteration because infeasibility occurs when solving the power system problem. Then, we enlarge the feasible intervals of bus voltages and line flows. However, B1 still fails to converge, and oscillation occurs in the charging power demands $D_i^{T}$ and energy-sharing prices $\lambda_i$, as shown in Fig. \ref{fig:B1}. For example, based on the outcomes of $D_i^{T}$ at iteration 2 where $D_1^T>D_2^T>D_3^T>D_4^T$, iteration 3 first clears the power market with $\lambda_1=1.76$\$/kWh, $\lambda_2=4.56$\$/kWh, $\lambda_3$=0.43\$/kWh, $\lambda_4$=0.44\$/kWh. The resulting price changes cause the change of EVs' routing in the transportation system, which gives $D_3^T>D_4^T>D_1^T=D_2^T$. As a result, at iteration 4, the sharing prices return to its previous values, leading to oscillation.
For B2, the computational complexity becomes a challenge as the KKT conditions of the energy sharing problem produce numerous complementary slackness constraints and require a lot of binary variables to yield a solvable form. It fails to return a result in 1 hour, the target time interval of this studied problem.
Our proposed method solves the MILP without iterations and does not have a convergence problem.
TABLE \ref{tab:compare} summarizes the utility/cost of power system (PS) and transportation system (TS) under B3 and the proposed method. In B3, prosumer 4 does not participate in energy sharing. From the table, we can find that the proposed method has a better utility (cost) of PS (TS) than B3. It indicates the positive influence of energy sharing on the joint operation of power and transportation system.

\begin{table}[htbp]
  \centering
  \caption{Cost comparison between different methods (Unit:\$).}\label{tab:compare}
    \begin{tabular}{ccc}
    \hline
             & Utility of PS      & Cost of TS  \\
             & $\sum_i U_i(d_i)$  & $\sum_{a\in T_A} [\omega t_a x_{ag}+(\gamma^c_a E_b+\omega t_a) x_{ae}]$\\
    \hline
    B3       & 3561.86 & 124743.91  \\
    Proposed & 3613.77 & 124619.38 \\
    \hline
    \end{tabular}%
\end{table}%

\begin{figure}[!htbp]
  \centering
  \includegraphics[width=0.4\textwidth]{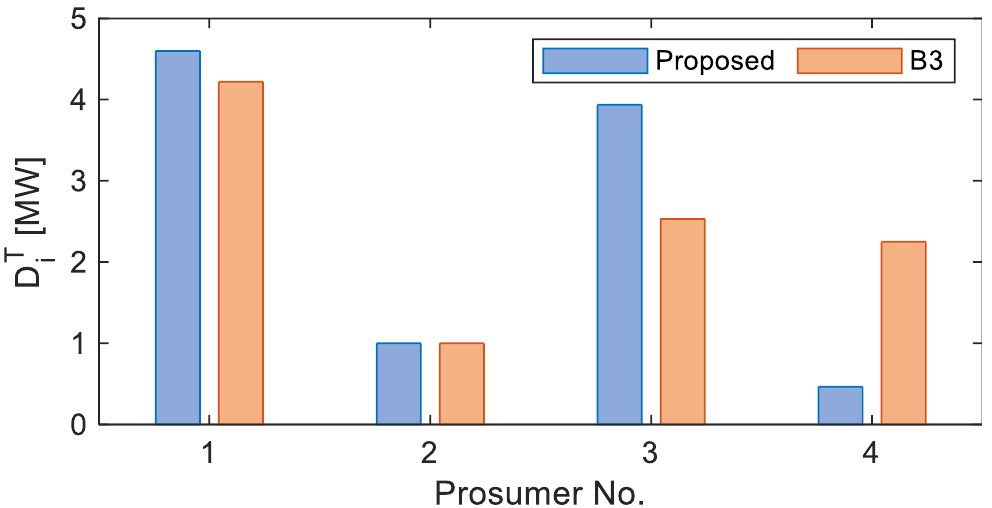}\\
  \caption{Illustration of the traffic flow of OD pair T1-T12 in B3 and the proposed method.}\label{fig:B3vsProp}
\end{figure}

We further compare the $\lambda_i$ and $D_i^{T}$ under B3 and the proposed method because the EVs' choice of CS is closely related to their offered charging prices.
Compared with B3, the sharing price difference between prosumer 3 and 4 under the proposed method is larger. This difference incentivizes some EVs to change their original routes determined by B3, and more EVs select charging stations operated by prosumer 3 that offers a relatively lower price, as shown in Fig. \ref{fig:B3vsProp}.
For example, for OD pair T4-T10, the proposed method's solution allocates all trip rate (0.5p.u.) to pass through E10-E12-E13-E14-E27-E28(prosumer 3)-E30. On the contrary, B3 dispatches a smaller EV trip rate (0.375p.u.) to go through that path and dispatch the remaining EV trip rate (0.125p.u.) to go through path E19-E21-E22-E32-E33(prosumer 4)-E35-E36 because of the close prices of prosumers 3 and 4.
This comparison suggests that energy sharing affects the EV traffic flow on the transportation system through pricing.

\begin{table}[htbp]
  \centering
  \caption{Sharing prices of B3 and proposed method (Unit: \$/kWh).}\label{tab:B3vsProp}%
    \begin{tabular}{ccccc}
    \hline
    Prosumer & 1     & 2     & 3     & 4 \\
    \hline
    B3    & 0.4392 & 0.4442 & 0.4431 & 0.4432 \\
    Proposed & 0.4336 & 0.4402 & 0.4342 & 0.4400 \\
    \hline
    \end{tabular}%
\end{table}%
\vspace{-0.cm}

\subsection{8-Prosumer System}
\begin{figure}[!htbp]
  \centering
  \includegraphics[width=0.4\textwidth]{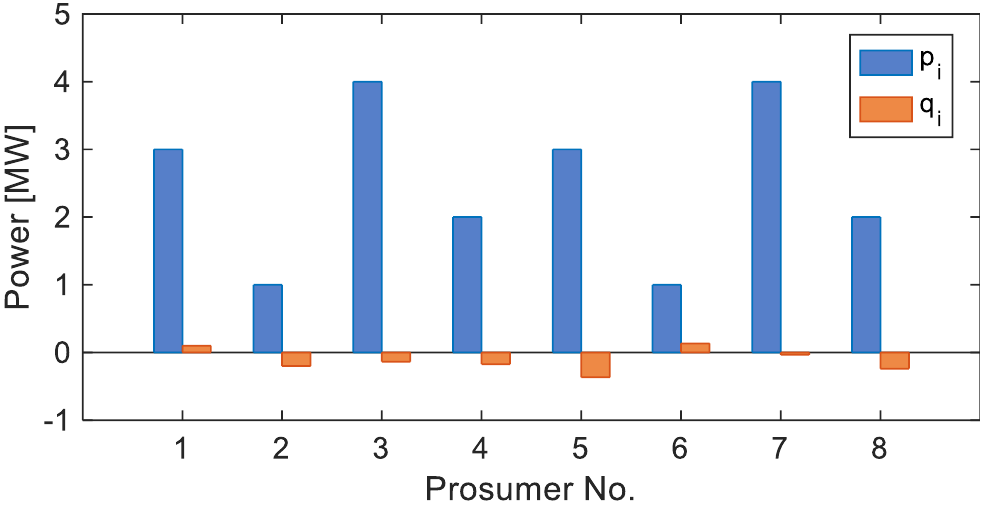}\\
  \caption{Energy production and sharing amount outcome of 8 prosumers.}\label{fig:8pro}
\end{figure}

The case studies and analyses above have demonstrated the effectiveness of the proposed method. Here, we increase the number of prosumers in the distribution network to 8 (Fig. \ref{fig:ieee33}), with each prosumer serving one CS in the transportation system (Fig. \ref{fig:TN}).
Fig. \ref{fig:8pro} shows the renewable generation and energy sharing amounts of each prosumer. In this energy sharing market, prosumers 1 and 6 become buyers and others are sellers, the roles of which are more diverse than the 4-prosumer system.

\begin{figure}[!htbp]
  \centering
  \includegraphics[width=0.4\textwidth]{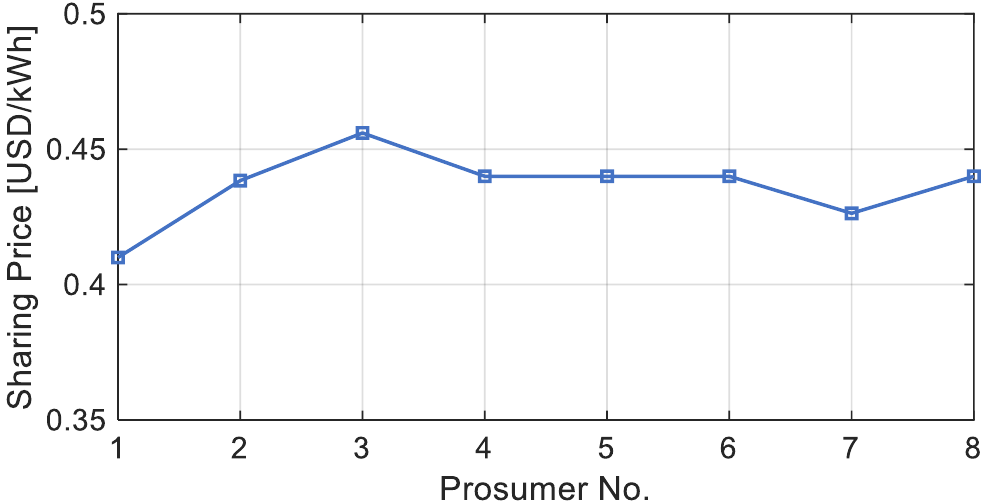}\\
  \caption{Sharing prices outcome of 8 prosumers.}\label{fig:8price}
\end{figure}

\begin{figure}[!htbp]
  \centering
  \includegraphics[width=0.4\textwidth]{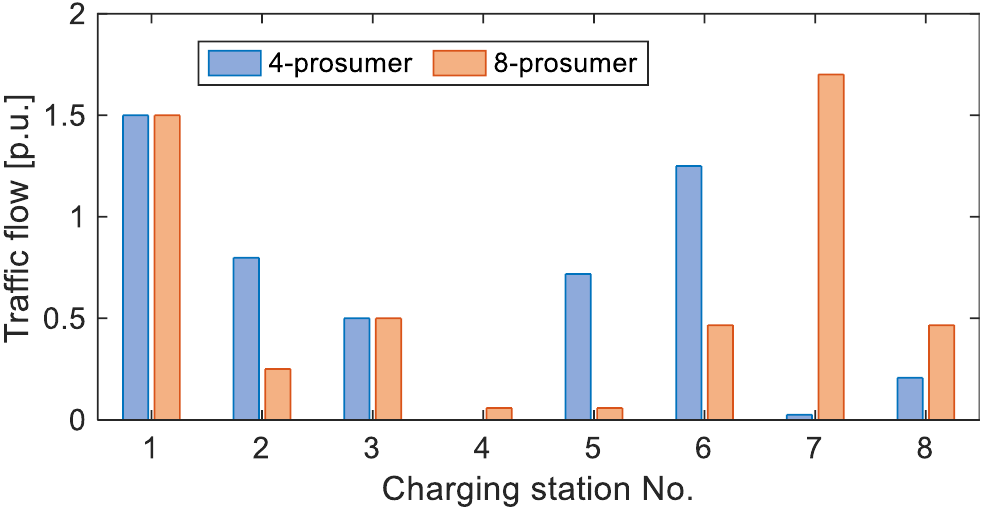}\\
  \caption{Traffic flow distribution of 4-prosumer and 8-prosumer system.}\label{fig:flow4vs20}
\end{figure}

The determined sharing prices are illustrated in Fig. \ref{fig:8price}. As shown, the sharing prices vary from 0.41 to 0.46\$/kWh. The lower prices appear in prosumers 1 and 7, which operate charging stations 1 and 7, respectively. Correspondingly, we can see in Fig. \ref{fig:flow4vs20} that the charging load of charging stations 1 and 7 are the highest. It reflects the impact of the sharing price on the EV routing. In addition, the computation time to solve this 8-prosumer case is 710s under the piecewise McCormick envelop of $|S|=3$, which is higher than the 4-prosumer system but is still acceptable.

\subsection{Impact of Parameters}
\subsubsection{Partition Number of Piecewise McCormick Envelope}
In this paper, the piecewise McCormick envelope technique is used to approximate the bilinear term. To investigate the impact of partition number $|S|$ on the accuracy, we change $|S|$ from $5$ to $20$. We calculate the error between the approximation value $\sigma$ in \eqref{eq:Mccormick} and the actual product of the values $D_i^{T}$ and $\lambda_i$, as shown in TABLE \ref{tab:partitionMcc}. Generally, a larger number of partitions used in the piecewise McCormick envelope generates a more accurate result. When $|S|=10$, the error has fallen within 5\%. As $|S|$ increases, the solving time increases significantly due to using more binary variables. Therefore, the choice of partition number depends on the trade-off between solution accuracy and computational efficiency.

\begin{table}[htbp]
  \centering
  \caption{The approximation error and solving time under different partitions.}  \label{tab:partitionMcc}%
    \begin{tabular}{cccccc}
    \hline
          & \multicolumn{1}{l}{Prosumer No. } & 1     & 2     & 3     & 4 \\
    \hline
    \multirow{2}{*}{$|S|$=5} & error & 4\%   & 18.18\% & 0.83\% & 18.18\% \\
    \cline{2-6}          & solver time & \multicolumn{4}{c}{80s} \\
    \hline
    \multirow{2}{*}{$|S|$=8} & error & 1\%   & 16.14\% & 16.28\% & 2.45\% \\
    \cline{2-6}          & solver time & \multicolumn{4}{c}{104s} \\
    \hline
    \multirow{2}{*}{$|S|$=10} & error & 1.49\% & 0.20\% & 1.33\% & 0\% \\
    \cline{2-6}          & solver time & \multicolumn{4}{c}{120s} \\
    \hline
    \multirow{2}{*}{$|S|$=20} & error & 0.38\% & 2.91\% & 2.33\% & 2.08\% \\
    \cline{2-6}          & solver time & \multicolumn{4}{c}{1644s} \\
    \hline
    \end{tabular}%
\end{table}%


\subsubsection{EV Charging Demand}
\begin{figure}[!htbp]
  \centering
  \includegraphics[width=0.4\textwidth]{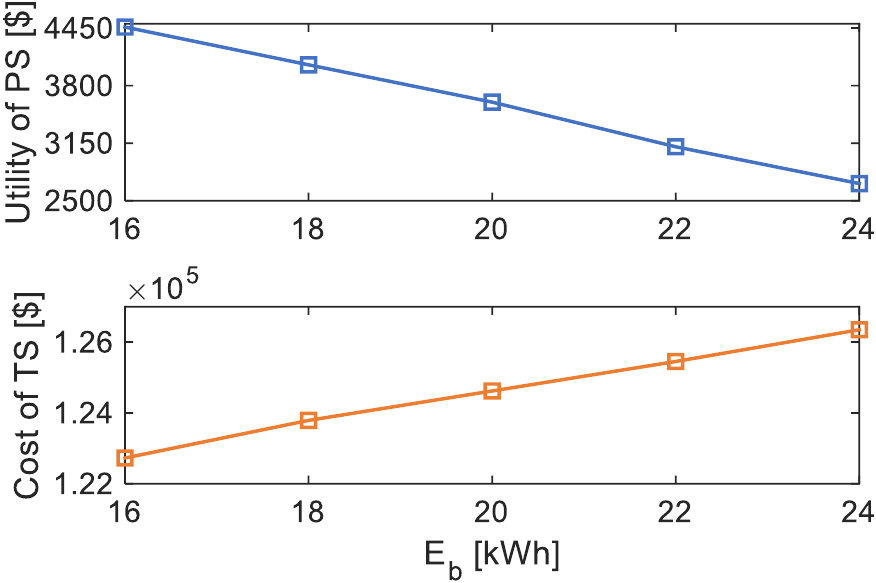}\\
  \caption{Impact of $E_b$ on the utility/cost of PS and TS.}\label{fig:Eb}
\end{figure}

Fig. \ref{fig:Eb} shows the impact of EV charging demand per car $E_b$ on the utility/cost of PS and TS. As the EV charging demand increases, the utility of PS decreases. This is because more electricity is required by the charging load, so less flexible load can be supplied and their utility decreases. In addition, as the EV charging demand increases, the cost of TS is not significantly affected. This indicates that EV charging demand has a major impact on the power system, while its influence on the transportation system is relatively small.

\subsubsection{Time Monetary Value}
\begin{figure}[!htbp]
  \centering
  \includegraphics[width=0.4\textwidth]{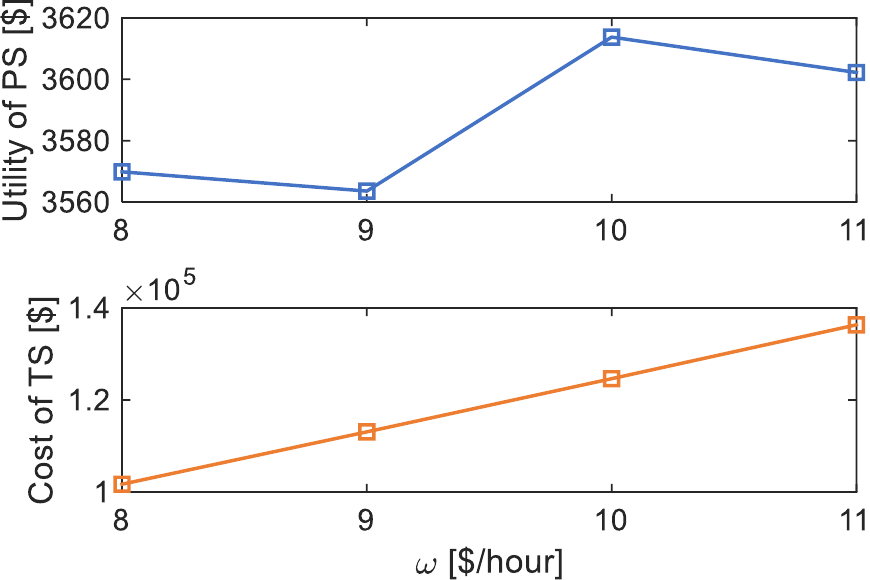}\\
  \caption{Utility/cost of PS and TS under different $\omega$.}\label{fig:w}
\end{figure}

Fig. \ref{fig:w} depicts the impact of the time monetary value $\omega$ on the utility/cost of PS and TS. As $\omega$ increases, the cost of TS significantly rises from $1.02\times 10^5$ to $1.36\times 10^5$, representing an increase of 33\%. In contrast, the utility of PS varies within a relatively narrow range of 3564\$ and 3614\$. This implies that the parameter $\omega$ mainly affects the transportation system and has little effect on power system.

{\color{black}
\subsubsection{EV Owner Preference}
In the transportation model, we assume EV users treat the travel cost and charging cost equally, denoted by Case0. In fact, different EV owners may have different preferences. To investigate this issue, we divide EV users into two groups: Group 1 comprises EV owners traveling between OD pairs T1-T6 and T1-T10, and Group 2 comprises others. We assume that the EV owners in Group 2 place more emphasis on the charging cost than the travel cost. By that, we multiply a coefficient 0.8 to the travel cost term, as denoted by Case1. The simulation results show that the EV routing and charging choices differ between Case0 and Case1. For example, in OD pair T1-T12, all EVs in Case0 select the path (E2-E3(CS1)-E5-E6-E27-E29-E30-E40), but in Case1 only 26\% of the EVs select this path while 74\% select the path (E1-E15-E16(CS3)-E18-E37-E41-E43-E44). This comparison demonstrates that the EV owner's preference can affect the routing and charging choices.
}

{\color{black}
\subsection{Test Case with IEEE 123-bus System}
To further validate the effectiveness of the proposed model and algorithm, the 33-bus power system is replaced with a larger one, a modified IEEE 123-bus system \cite{huang2022improved}. As shown in Fig. \ref{fig:123bus}, there are 16 prosumers (P1-P16) located at different buses. Specifically, prosumers P1-P8, which are located at buses 14, 43, 73, 102, 26, 61, 97, and 115, respectively, operate charging stations CS1-CS8.
The partition number is set to 5 in the piecewise McCormick envelope. Due to the larger scale of the coupled system, it takes a longer time, 464 seconds, to compute the game equilibrium of the coupled power-transportation system. 
At the equilibrium, the settled energy sharing prices and shared energy quantities of prosumers are shown in Fig. \ref{fig:lbdi_qi}. We can find that P2, P5, and P7 are buyers while the other prosumers are sellers.

Meanwhile, the traffic flow and optimal EV routing and charging decisions are determined in the transportation system. Taking OD pair T1-T12 as an example, all EVs select the most cost-effective path (E1-E15-E16(CS3)-E18-E37-E41-E43-E44), and charge at CS3 in road E16. The incurred cost is 17.97\$. If EVs select other paths, such as E7-E10-E11-E13-E23-E25-E26-E39, a higher cost of 25.19\$ will be incurred. Overall, the results validate that the proposed model and algorithm are applicable to the larger IEEE 123-bus system.

\begin{figure}[!htbp]
  \centering
  \includegraphics[width=0.45\textwidth]{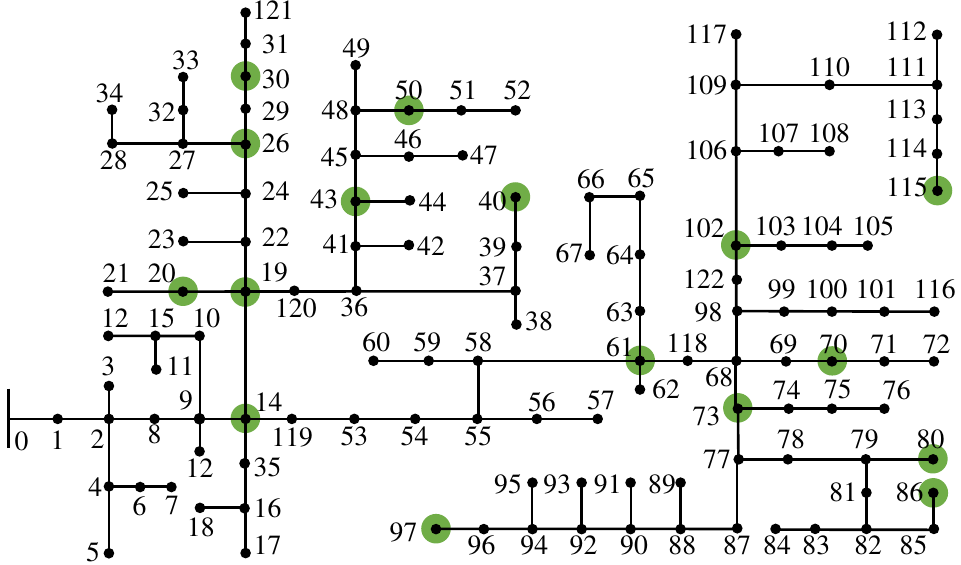}\\
  \caption{{\color{black}Modified IEEE 123-bus test system with 16 prosumers marked in green.}}\label{fig:123bus}
\end{figure}

\begin{figure}[!htbp]
  \centering
  \includegraphics[width=0.4\textwidth]{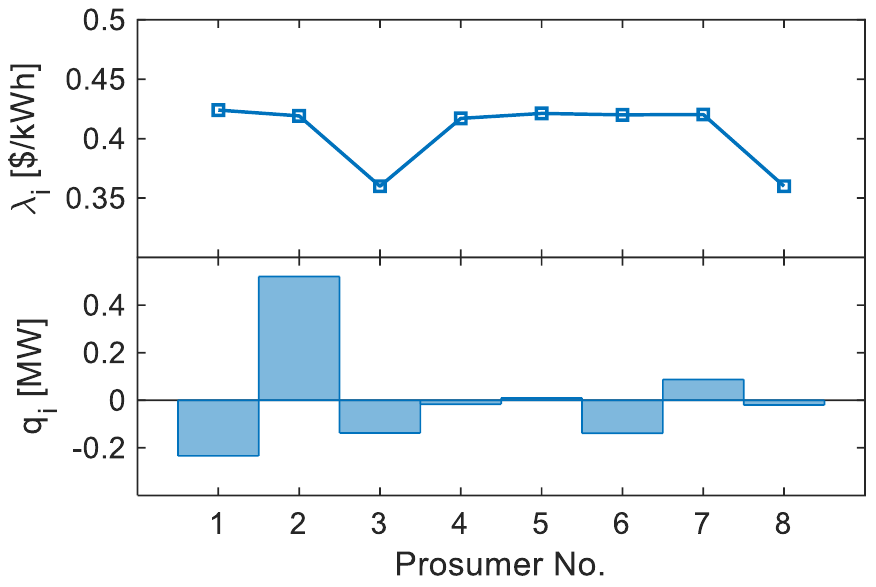}\\
  \caption{{\color{black}Sharing prices $\lambda_i$ and shared energy quantities $q_i$ of prosumers P1-P8.}}\label{fig:lbdi_qi}
\end{figure}

\subsection{Computation Time}
For the 4-prosumer case in the IEEE 33-bus power grid, the computing time is 120 seconds, much shorter than the time interval of 1 hour. In the 8-prosumer system case, computing the equilibrium takes 710 seconds. For the 16-prosumer case in the IEEE 123-bus power grid, computing the equilibrium takes 464 seconds. Generally, the computation time grows when the system becomes larger and more complex, while also being affected by the network topology and the load distribution across nodes. Overall, the proposed method is still viable for the real-time (single period: 1 hour) management of the coupled power and transportation system.
}

\section{Conclusion}
This paper presents a novel hierarchical game model to characterize the complex interaction between transportation and power system. In the power system, we propose an energy sharing mechanism for prosumers, which casts down to a generalized Nash game. In the transportation system, all drivers strategically choose their routes with the goal of minimizing their travel costs, resulting in a Nash game. Externally, the two systems constitute a Nash game.
To analyze the equilibrium of this hierarchical game, two equivalent models are provided. Then optimality conditions and linearization techniques are employed to convert the hierarchical game model into a MILP, enabling efficient computation.
Case studies demonstrate the effectiveness and benefits of the proposed method. We have the following findings:
1) The EV routing to charging stations is closely related to their offering prices.
2) Prosumers engaging in energy sharing can reduce the operation costs of both systems.
3) The proposed method can greatly reduce the computational time and avoid the non-convergence issue of the traditional best response decomposition method. 

{\color{black}Future research may consider the temporal evolution of EV flows and power demands and extend the scheduling of coupled power and transportation systems to a multi-period scenario. Moreover, in addition to energy sharing in the power system, ride sharing in the transportation system will also be considered.}

\ifCLASSOPTIONcaptionsoff
\newpage
\fi

\bibliographystyle{IEEEtran}
\bibliography{IEEEabrv,mybib}

  
\appendices
\makeatletter
\@addtoreset{equation}{section}
\@addtoreset{theorem}{section}
\makeatother
\setcounter{equation}{0}  
\renewcommand{\theequation}{A.\arabic{equation}}
\renewcommand{\thetheorem}{A.\arabic{theorem}}
\section{Proof of Proposition \ref{prop-1}}
\label{apendix-A}
Denote $q_i=-m_a\lambda_i+b_i,\forall i\in \mathcal{E}_N$. 
Then problem \eqref{eq:market} can be equivalently written as
\bsq
\begin{align}
    \mathop{\min}_{q_i,\forall i} ~ & \sum_{i \in \mathcal{E}_N} (q_i-b_i)^2  \\
    \mbox{s.t.}~ & q \in \mathcal{F}_c
\end{align}
\esq
Suppose $(b^*,d^*,\lambda^*)$ is the GNE of the energy sharing game \eqref{eq:market} and \eqref{eq:eachnode}, then for problem \eqref{eq:market} we have
\begin{align}
\label{eq:condition-3}
		\sum \limits_{i \in \mathcal{E}_N} (q_i-q_i^*)(q_i^*-b_i^*) \ge 0,\forall q \in \mathcal{F}_c
\end{align}
Let $\mathcal{D}_i:=\{d_i~|~ \underline{D}_i \le d_i \le \overline{D}_i\}$. For problem \eqref{eq:eachnode}, first we substitute \eqref{eq:eachnode.2} into the objective function \eqref{eq:eachnode.1} to eliminate $b_i$, then for all $i\in\mathcal{E}_N$ its optimality condition is
\begin{align}
\label{eq:condition-2}
		-U_i(d_i)+U_i(d_i^*)+(d_i-d_i^*)\lambda_i^*\ge 0, \forall d \in \mathcal{D}_i
\end{align}

For problem \eqref{eq:share-eq}, its Lagrangian function defined on $\Omega:=\prod_{i \in \mathcal{E}_N} \mathcal{D}_i \times \mathcal{F}_c \times \mathbb{R}^{|\mathcal{E}_N|}$ is
\begin{align}
L(d,q,\xi)=~&\sum \limits_{i \in \mathcal{E}_N} -U_i(d_i) +\sum \limits_{i \in \mathcal{E}_N} \xi_i(d_i+d_i^f+D_i^{T}-p_i-q_i)
\end{align}
Let $(\hat{d},\hat q,\hat{\xi})$ be a saddle point of the Lagrangian function, then $(\hat d, \hat q, \hat \xi) \in \Omega$ and it satisfies $\forall (d,q,\xi)\in \Omega$:
\bsq
\label{eq:condition}
\begin{align}
\left[-U_i(d_i)+U_i(\hat d_i)+ (d_i-\hat d_i)\hat \xi_i\right] & ~\ge 0 ,\forall i \in \mathcal{E}_N \label{eq:condition.1}\\
-\sum \limits_{i \in \mathcal{E}_N} (q_i-\hat q_i)(\hat \xi_i) &~\ge 0 \label{eq:condition.2}\\
\sum \limits_{i \in \mathcal{E}_N}  (\xi_i-\hat \xi_i)(\hat d_i+d_i^f+D_i^{T}-p_i-\hat q_i) & ~\le 0\label{eq:condition.3}
\end{align}
\esq

\emph{\textbf{Existence}}. If problem \eqref{eq:share-eq} is feasible, suppose $\hat d$ is its optimal solution and $\hat \xi$ is the corresponding dual variable. Let $d^*=\hat d$, $\lambda^*=\hat \xi$, $q_i^*=\hat d_i+d_i^f+D_i^{T}-p_i$ for all $i \in\mathcal{E}_N$, and $b_i^*=m_a\hat \xi_i+q^{*}_i,\forall i \in \mathcal{E}_N$, then it is easy to check that \eqref{eq:condition-2} and \eqref{eq:condition-3} are met. Thus, we have constructed a GNE $(d^*,b^*,\lambda^*)$.
		
\emph{\textbf{Uniqueness}}. Given a GNE $(d^*,b^*,\lambda^*)$, we have $b_i^*=d_i^*+d_i^f+D_i^{T}-p_i+m_a\lambda_i^*$ for all $i \in \mathcal{E}_N$. Let $\hat d=d^*$, $\hat \xi=\lambda^*$, and $\hat q=-m_a\lambda^*+b^*$, then it is easy to check that $(\hat d,\hat q,\hat \xi)$ satisfies \eqref{eq:condition}, so $\hat d$ is the optimal solution of \eqref{eq:share-eq} and $\hat \xi$ is the corresponding dual optimum. Since the objective function is strictly convex, and the constraint sets $\mathcal{D}_i,\forall i \in \mathcal{E}_N$  and $\mathcal{F}_c$ are all compact convex sets, problem \eqref{eq:share-eq} attains a unique solution, so $\hat d$ is unique and problem \eqref{eq:share-eq} is feasible.

\setcounter{equation}{0}  
\renewcommand{\theequation}{B.\arabic{equation}}
\renewcommand{\thetheorem}{B.\arabic{theorem}}
\section{Proof of Proposition \ref{prop-2}}
\label{apendix-B}
The Lagrangian of the equivalent optimization problem \eqref{eq:traffic-eq} with respect to the equality constraints \eqref{eq:traffic-eq.5} and \eqref{eq:traffic-eq.6} is
\begin{align}
    \mathcal{L}(f_{kg}^{rs},f_{ke}^{rs},u_{g}^{rs},u_{e}^{rs})= \underbrace{\omega \sum_{a \in \mathcal{T}_A} \int_0^{x_a} t_a(\theta) d\theta + \sum_{a\in T_A^C} \gamma_a^c E_b x_a}_{\mathcal{L}_1} \nonumber\\
    + \underbrace{\sum_{rs} u^{rs}_g\left(q^{rs}_g-\sum_{k \in \mathcal{K}_g^{rs}}f_{kg}^{rs}\right)}_{\mathcal{L}_2}
    + \underbrace{\sum_{rs} u^{rs}_e\left(q^{rs}_e-\sum_{k \in \mathcal{K}_e^{rs}}f_{ke}^{rs}\right)}_{\mathcal{L}_3}
\end{align}
Therefore, the KKT condition is
\bsq
\label{eq:KKT}
\begin{align}
    ~ & f_{kg}^{rs} \ge 0,\forall k \in \mathcal{K}_g^{rs}, \forall rs \label{eq:KKT-1}\\
    ~ & f_{kg}^{rs} \frac{\partial \mathcal{L}}{\partial f_{kg}^{rs}} =0, \forall k \in \mathcal{K}_{g}^{rs}, \forall rs \label{eq:KKT-2}\\
    ~ & \frac{\partial \mathcal{L}}{\partial f_{kg}^{rs}} \ge 0, \forall k \in \mathcal{K}_g^{rs}, \forall rs \label{eq:KKT-3} \\
    ~ & \sum_{k \in \mathcal{K}_g^{rs}} f_{kg}^{rs} = q_g^{rs},\forall rs \label{eq:KKT-4} \\
    ~ & f_{ke}^{rs} \ge 0,\forall k \in \mathcal{K}_e^{rs}, \forall rs \label{eq:KKT-5}\\
    ~ & f_{ke}^{rs} \frac{\partial \mathcal{L}}{\partial f_{ke}^{rs}} =0, \forall k \in \mathcal{K}_{e}^{rs}, \forall rs \label{eq:KKT-6}\\
    ~ & \frac{\partial \mathcal{L}}{\partial f_{ke}^{rs}} \ge 0, \forall k \in \mathcal{K}_e^{rs}, \forall rs \label{eq:KKT-7}\\
    ~ & \sum_{k \in \mathcal{K}_e^{rs}} f_{ke}^{rs} = q_e^{rs},\forall rs \label{eq:KKT-8}
\end{align}
\esq
The value of $\frac{\partial \mathcal{L}}{\partial f_{kg}^{rs}}$ can be calculated as follows:
\begin{align}
    \frac{\partial \mathcal{L}_1}{\partial f_{kg}^{rs}} = ~ & \frac{\partial \left(\omega \sum_{a \in \mathcal{T}_A} \int_0^{x_a}  t_a(\theta) d\theta\right)}{\partial x_a} \frac{\partial x_a}{\partial f_{kg}^{rs}} \nonumber\\
    = ~ & \sum_{a \in \mathcal{T}_A} \omega t_a(x_a)  \delta_{akg}^{rs} = c_{kg}^{rs}
\end{align}
Moreover,
\begin{align}
    \frac{\partial \mathcal{L}_2}{\partial f_{kg}^{rs}} = ~ & \frac{\partial \sum_{rs} u_g^{rs} \left(q_g^{rs}-\sum_{k \in \mathcal{K}_g^{rs}}f_{kg}^{rs}\right)}{\partial f_{kg}^{rs}}=-u_g^{rs}
\end{align}
Thus, the KKT conditions \eqref{eq:KKT-1}-\eqref{eq:KKT-4} are equivalent to
\bsq \label{eq:KKTeq1}
\begin{align}
    ~ & f_{kg}^{rs} \ge 0,\forall k \in \mathcal{K}_g^{rs}, \forall rs\\
    ~ & f_{kg}^{rs} (c_{kg}^{rs}-u_{g}^{rs}) =0, \forall k \in \mathcal{K}_g^{rs}, \forall rs\\
    ~ & c_{kg}^{rs}-u_{g}^{rs} \ge 0, \forall k \in \mathcal{K}_g^{rs}, \forall rs \\
    ~ & \sum_{k \in \mathcal{K}_g^{rs}} f_{kg}^{rs} = q_g^{rs},\forall rs
\end{align}
\esq
Similarly, the value of $\frac{\partial \mathcal{L}}{\partial f_{ke}^{rs}}$ can be calculated as follows:
\begin{align}
    \frac{\partial \mathcal{L}_1}{\partial f_{ke}^{rs}} = ~ & \frac{\partial \left(\omega \sum_{a \in \mathcal{T}_A} \int_0^{x_a}  t_a(\theta) d\theta+\sum_{a\in T_A^C} \gamma_a x_a E_b\right)}{\partial x_a} \frac{\partial x_a}{\partial f_{ke}^{rs}} \nonumber\\
    = ~ & \sum_{a \in \mathcal{T}_A} \omega t_a(x_a) \delta_{ake}^{rs} + \sum_{a \in T_A^C} \gamma_a^c E_b \delta_{ake}^{rs} = c_{ke}^{rs}
\end{align}
where the second equation is due to \eqref{eq:x-f}, and the last equation is according to the definition of $c_{ke}^{rs}$. Moreover,
\begin{align}
    \frac{\partial \mathcal{L}_3}{\partial f_{ke}^{rs}} = ~ & \frac{\partial \sum_{rs} u_e^{rs} \left(q_e^{rs}-\sum_{k \in \mathcal{K}_e^{rs}}f_{ke}^{rs}\right)}{\partial f_{ke}^{rs}}=-u_e^{rs}
\end{align}
Thus, the KKT conditions \eqref{eq:KKT-5}-\eqref{eq:KKT-8} are equivalent to
\bsq\label{eq:KKTeq2}
\begin{align}
    ~ & f_{ke}^{rs} \ge 0,\forall k \in \mathcal{K}_e^{rs}, \forall rs\\
    ~ & f_{ke}^{rs} (c_{ke}^{rs}-u_{e}^{rs}) =0, \forall k \in \mathcal{K}_e^{rs}, \forall rs\\
    ~ & c_{ke}^{rs}-u_{e}^{rs} \ge 0, \forall k \in \mathcal{K}_e^{rs}, \forall rs \\
    ~ & \sum_{k \in \mathcal{K}_e^{rs}} f_{ke}^{rs} = q_e^{rs},\forall rs.
\end{align}
\esq
The \eqref{eq:KKTeq1} and \eqref{eq:KKTeq2} are exactly the condition for Wardrop User Equilibrium. Furthermore, since $d t_a(x_a)/d {x_a}>0$ and $d t_a(x_a)/d x_b=0$, then $\frac{\partial^2 \mathcal{L}_1}{\partial x_a^2}=\frac{ \omega \partial t_a(x_a)}{\partial x_a} >0$, $\frac{\partial^2 \mathcal{L}_1}{\partial x_a \partial x_b} = \frac{\omega \partial t_a(x_a)}{\partial x_b}=0$. So the Hessian Matrix of $\mathcal{L}_1$ is a diagonal matrix with the diagonal elements all greater than 0, and is positive definite. The objective function of \eqref{eq:traffic-eq} is strictly convex and its feasible region \eqref{eq:traffic-eq.2}-\eqref{eq:traffic-eq.6} is also convex. Consequently, problem \eqref{eq:traffic-eq} has an optimal solution, where $x_a,\forall a \in \mathcal{T}_A$ is unique.

\end{document}